\documentclass[pre,showpacs,showkeys,nofootinbib,superscriptaddress,twocolumn]{revtex4-2}

\usepackage{amsmath,amssymb}
\usepackage{graphicx}
\usepackage[breaklinks=true,colorlinks=true,linkcolor=blue,urlcolor=blue,citecolor=blue]{hyperref}
\usepackage{mathtools}
\usepackage{url}

\def\Var{\mathrm{Var}}

\def\ve{\varepsilon}

\def\P{\mathbb{P}}

\def\R{\mathbb{R}}

\def\T{\mathcal{T}}

\def\erf{\mathrm{erf}}
\def\erfc{\mathrm{erfc}}

\def\ttau{\bar{\tau}}

\begin{document}

\title{Fastest first-passage time statistics for time-dependent particle
injection}

\author{Denis~S.~Grebenkov}
\email{denis.grebenkov@polytechnique.edu}
\affiliation{Laboratoire de Physique de la Mati\`{e}re Condens\'{e}e,\\ 
CNRS -- Ecole Polytechnique, Institut Polytechnique de Paris, 91120
Palaiseau, France}
\author{Ralf Metzler}
\affiliation{Institute for Physics \& Astronomy, University of Potsdam, 14476
Potsdam-Golm, Germany}
\affiliation{Asia Pacific Center for Theoretical Physics, Pohang 37673, Republic
of Korea}
\author{Gleb Oshanin}
\affiliation{Sorbonne Universit\'e, CNRS, Laboratoire de Physique Th\'eorique
de la Mati\`ere Condens\'ee (UMR CNRS 7600), 4 Place Jussieu, 75252 Paris
Cedex 05, France}
\affiliation{Asia Pacific Center for Theoretical Physics, Pohang 37673, Republic
of Korea}

\date{\today}

\begin{abstract}
A common scenario in a variety of biological systems is that multiple particles
are searching in parallel for an immobile target located in a bounded domain,
and the fastest among them that arrives to the target first triggers a given
desirable or detrimental process. The statistics of such extreme events---the
\textit{fastest\/} first-passage to the target---is well-understood by now
through a series of theoretical analyses, but exclusively under the
assumption that all $N$ particles start \textit{simultaneously\/}, i.e., all
are introduced into the domain instantly, by $\delta$-function-like pulses.
However, in many practically important situations this is not the case: in
order to start their search, the particles often have to enter first into a
bounded domain, e.g., a cell or its nucleus, penetrating through gated channels
or nuclear pores. This entrance process has a random duration so that the
particles appear in the domain sequentially and with a time delay. Here we
focus on the effect of such an extended-in-time injection of multiple
particles on the fastest first-passage time (fFPT) and its statistics. We
derive the full probability density function $H_N(t)$ of the fFPT with an
arbitrary time-dependent injection intensity of $N$ particles. Under rather
general assumptions on the survival probability of a single particle and on
the injection intensity, we derive the large-$N$ asymptotic formula for the
mean fFPT, which is quite different from that obtained for the instantaneous
$\delta$-pulse injection. The extended injection is also shown to considerably
slow down the convergence of $H_N(t)$ to its large-$N$ limit---the Gumbel
distribution---so that the latter may be inapplicable in the most relevant
settings with few tens to few thousands of particles.
\end{abstract}



\maketitle

\section{Introduction}

"The winner takes it all"-situation in which the fastest out of many actors
produces a required action is realized in diverse processes in physiology,
neuroscience, and cellular biology. For instance, 300 millions of motile
sperms search for the egg cell, and the first among them joins an ovum to
form a zygote \cite{Reynaud15}. Many thousands of cells (e.g., bacteria) in
a colony compete with each other to first respond to a common environmental
challenge \cite{Klinger16,Flemming19}. In cellular processes, a large amount
of messengers speed up and control signal transduction and cell
communication; e.g., the fastest among several hundreds of calcium ions
injected into a dendritic spine triggers a transduction, while several
thousands of neurotransmitters diffusing in the pre-synaptic terminal search
for receptors on the post-synaptic membrane \cite{Fain,Hille}. In the cell
nucleus, many transcription factors seek in parallel a specific binding site
on the cellular DNA \cite{Alberts}. Further examples can be found in a recent
monograph \cite{Denis}.

In the mathematical modeling of these processes (see, e.g.,
\cite{Weiss83,Meerson15,Schuss19,Basnayake19,Lawley20,Lawley20b,Lawley20c,Madrid20,Grebenkov20a,Grebenkov22,LawleyBook}),
a common theme is to suppose that all particles are injected into the
domain simultaneously, by $\delta$-function-like pulses.  This is
tantamount to the tacit assumption that the duration of the injection
process is much shorter than the typical time-scales of the subsequent
search processes---certainly a plausible scenario in the case of the
sperm cells, but not in general. In particular, transcription factors
in cells typically do not start at the same time instant but are
produced in intermittent bursts and/or at different locations
\cite{Alberts,otto,xie} as well as do not spread evenly over the cell
\cite{kuhlman,joel}. In the case of viral infections, the viruses may
keep on entering a single living cell, or injecting their RNA/DNA into
it, within significantly extended periods of time \cite{ptashne}.
Similarly, the injection of ions into a cell through gated channels is
typically spread over time \cite{zilman1,zilman2}.  Such
channels---molecular machines (see Fig.~\ref{fig:channel})
\cite{Hille}, which transport ions through the cell membrane with high
efficiency of $10^6-10^8$ ions/s---are typically open for a few
milliseconds.  As the diffusion coefficient of an ion in the cytoplasm
is typically of the order of a few tens of ${\rm\mu m^2/s}$, the first
ion entering the cell may diffuse away from the point of injection on
quite noticeable distances of a few hundreds of nanometers, before the
last ion in the pulse even enters the cell. Since the typical reaction
times themselves are often of order of milliseconds, or even much
shorter---as in case of diffusion-controlled reaction of a single
calcium ion to a calcium-sensor protein
\cite{Hodgkin52,Sheng12,Nakamura15,Brockhaus19,Reva21}---the reaction
event may take place well before the last ion enters the cell. In
consequence, only some portion of the injected ions will effectively
contribute to the search process such that the fastest first-passage
time (fFPT) to the reaction event and other characteristic time
scales, calculated using the assumption of an injection via a
$\delta$-pulse, may considerably underestimate the actual
characteristics of these properties and therefore present an
inadequate picture of the kinetic behavior.  Moreover, one can
even aim at optimizing the search process by adjusting the injection
rate or injecting individual searchers at prescribed time instances
\cite{Campos24,Meyer25}.

Here, we combine analytical and numerical tools to address the
conceptually important question on how the duration of the injection
stage or, more specifically, the time dependence of the injection
intensity $\psi(t)$ affects the statistics of the fFPT to the reaction
event---and how different it can be from the one corresponding to the
situation when all $N$ particles are simultaneously injected to the
reaction bath by a $\delta$-pulse. We derive exact expressions for the
probability density function (PDF) of the fFPT for a general
time-dependent injection profile $\psi(t)$. Next, we analyze how the
form of $\psi(t)$ affects the asymptotic large-$N$ behavior of the
mean fFPT and of the full fFPT-PDF. In particular, we demonstrate that
an extended-in-time injection drastically slows down the convergence
of the fFPT-PDF to the Gumbel distribution, which is the fingerprint
feature of the $\delta$-pulse injection
\cite{Lawley20,Lawley20b,Lawley20c,Madrid20,Grebenkov20a,LawleyBook}.

The paper is outlined as follows. In Sec.~\ref{sec:1} we formulate our
model and introduce general notations as well as define two basic
injection scenarios.  Section \ref{sec:2} contains our theoretical
results. In particular, Sec.~\ref{sec:3} presents the large-$N$
asymptotic behavior of the mean fFPT for an extended-in-time particle
injection and discusses three possible regimes according to the
injection profile. Section \ref{sec:4} is dedicated to a numerical
validation of our predictions for a particular case of one-dimensional
diffusion. Finally, in Sec.~\ref{sec:5} we conclude with a brief
summary of our results. Details of intermediate calculations are
relegated to Appendices.

\section{Model and notations} 
\label{sec:1}

\begin{figure}
\begin{center}
\includegraphics[width=85mm]{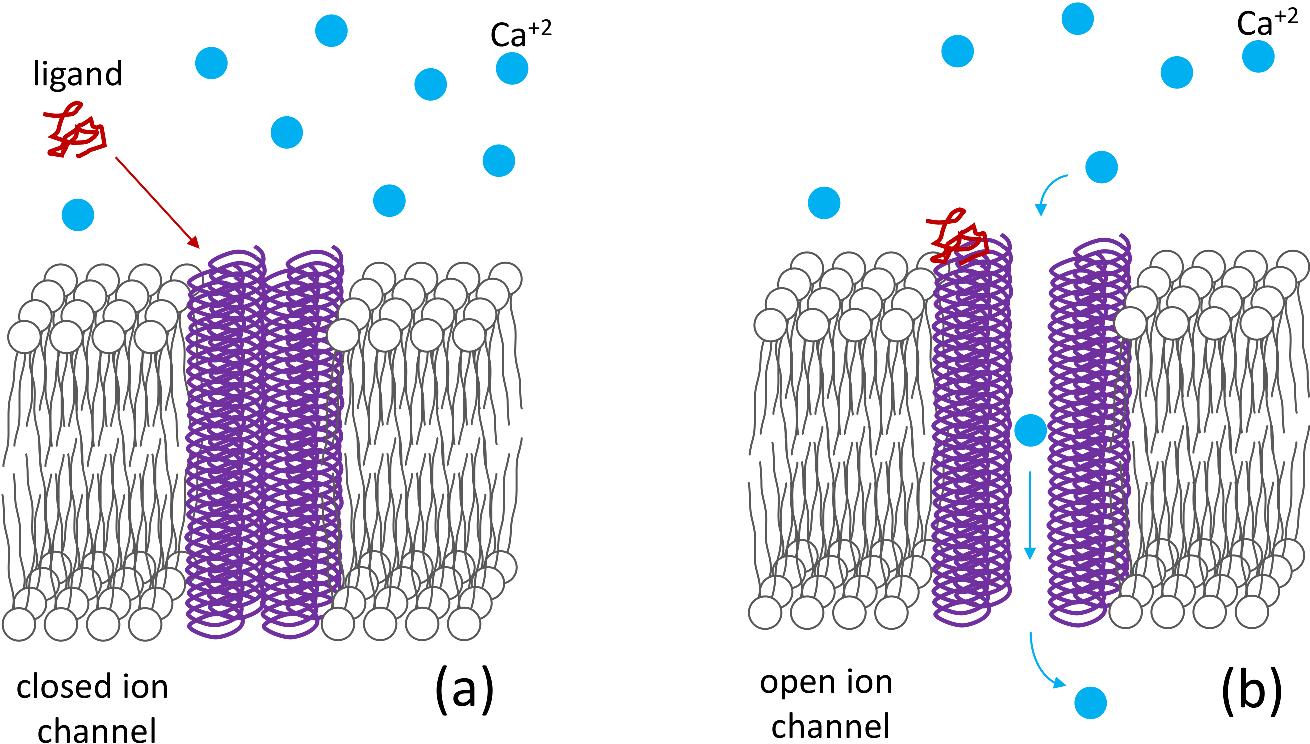} 
\end{center}
\caption{Illustration of a sequential influx of calcium ions into a cell
through an ion channel. In panel (a) the channel is closed while panel (b)
depicts a channel, which opens due to the arrival of a ligand to a special
binding site \cite{Hille}.}
\label{fig:channel}
\end{figure}

Consider a bounded domain in which $N$ particles are sequentially injected at
some fixed time instants $\delta_1,\delta_2,\ldots,\delta_N$. Note that all
particles may appear in the domain at the same location, e.g., corresponding
to the passage through a single ion channel (see Fig.~\ref{fig:channel}), as
it was assumed in \cite{Lawley20,Lawley20b,Lawley20c,Madrid20,Grebenkov20a,
Grebenkov22,LawleyBook}. Alternatively, the starting points can be drawn from
a given spatial distribution, e.g., it can be uniformly distributed on the
plasma membrane as it happens when viruses are penetrating into the cell
\cite{a,ptashne}. All derivations below remain valid for both cases, but
we stress that the behavior of the fFPT may strongly depend on the precise
spatial distribution of the starting points.

We assume that the domain contains a target site which the $N$ particles
are searching for \textit{independently\/} of each other. Let $\tau_k+
\delta_k$ denote the random time instant when the $k$th particle reaches
the target for the first time and let $S(t-\delta_k)$ (with the convention
$S(t\leq 0)\equiv1$) denote the corresponding "survival" probability, i.e.,
the probability that the $k$th particle did not arrive to the target within
the time interval $t-\delta_k$. Note that since the domain is bounded, one
evidently has $S(t)\to0$ as $t\to\infty$, such that any particle is
\textit{certain\/} to find the target.  Then, denoting the fFPT as
\begin{equation}
\label{fFPT}
\T_N=\min\{\tau_1+\delta_1,\ldots,\tau_N+\delta_N\}, 
\end{equation}
one writes the probability that none of the $N$ particles has reached the
target up to time $t$ as
\begin{equation}
\label{z}
\P\{\T_N>t\}=\prod\limits_{k=1}^NS(t-\delta_k),
\end{equation}
for a given set of time instants $\delta_1,\delta_2,\ldots,\delta_N$. In what
follows, we consider two possible scenarios with respect to the variables
$\delta_k$:

A. {\it Deterministic injection.}  In this scenario we assume that a fraction
$N\psi(t)dt$ of particles enters the domain within the time interval $(t,t+dt)$,
with a given flux profile $\psi(t)$ per particle.

B. {\it Random injection.} Here, we suppose that $\delta_k$ are independent,
identically-distributed random variables with common probability density
$\psi(t)$.

For both scenarios, the survival probability $S_N(t)$ of the target in the
presence of $N$ particles with a time-extended injection is simply obtained
by averaging the probability $\P\{\T_N>t\}$ with respect to the distribution
of the variables $\delta_k$. Correspondingly, the PDF $H_N(t)$ of the fFPT
is the time-derivative of $S_N(t)$, taken with the negative sign,
\begin{equation}  
\label{HH}
H_N(t)=-\frac{dS_N(t)}{dt},
\end{equation}
while the mean fFPT obeys
\begin{equation}
\label{meanfFPT}
\langle\T_N\rangle=\int^{\infty}_0dt\,t\,H_N(t)=\int^{\infty}_0dt\,S_N(t).
\end{equation}
Note, as well, that for situations in which the particles start from random
positions, the survival probability $S(t)$ is supposed to already include
this spatial average.

\section{Analytical results} 
\label{sec:2}

\subsection{Deterministic injection}

For the deterministic injection scenario, the formal expression
\eqref{z} is neither suitable for the asymptotic analysis for large
$N$, nor for the analytical inspection of the role of the entrance
delays. To overcome this limitation, we describe the non-instantaneous
injection of particles by introducing a piecewise-constant function
$N(t)$ that increases by unity at each injection event and counts the
number of particles that have entered up to time $t$. When $N$ is
large, it is more convenient to "smooth" the rescaled profile
$\eta(t)=N(t)/N$, i.e., to consider it as a smooth function of time
which increases from zero at $t=0$ to unity in the limit
$t\to\infty$. We can therefore define the influx of particles at time
$t$ as $\psi(t)=d\eta(t)/dt$, so that $N\psi(t)\delta$ is the
"number" of particles having entered the domain during a short time
period from $t$ to $t+\delta$. In doing so, one finds
\begin{eqnarray} 
\nonumber
\langle\P\{\T_N>t\}\rangle_{\delta_k}&\approx&[S(t)]^{N\psi(0)\delta}\,[
S(t-\delta)]^{N\psi(\delta)\delta}\ldots\\
&&\times[S(0)]^{N\psi(t)\delta},
\end{eqnarray}
where the brackets denote averaging over the entrance times $\delta_1,\ldots,
\delta_N$. In the continuous limit $\delta\to0$, the product on the right-hand
side converges to the expression
\begin{equation}
\langle\P\{\T_N>t\}\rangle_{\delta_k}\xrightarrow[\delta\to0]{}\bar{S}_N(t)= [\bar{S}
_\psi(t)]^N,
\end{equation}
where $\bar{S}_N(t)$ denotes the survival probability of the
target in the presence of $N$ particles injected into the domain with
the smooth time-dependent injection profile $\psi(t)$, and
\begin{equation}
\label{Spsibar}
\bar{S}_\psi(t)=\exp\left(\int\limits_0^tdt'\,\psi(t')\,\ln S(t-t')\right).
\end{equation}
As a consequence, we find for the deterministic injection that
\begin{equation}
\label{eq:Spsi2}
\bar{S}_N(t) =\exp\left(N\int\limits_0^tdt'\,\psi(t')\,\ln S(t-t')\right).
\end{equation}
Respectively, in virtue of Eq.~\eqref{HH}, the fFPT-PDF obeys
\begin{equation}  
\label{Hbar}
\bar{H}_N(t)=-\frac{d}{dt}[\bar{S}_\psi(t)]^N, 
\end{equation}
while the mean fFPT, by virtue of Eq.~\eqref{meanfFPT}, is determined by the
integral
\begin{equation}
\label{mean1bar}
\langle\bar{\T}_N\rangle=\int^{\infty}_0dt\,[\bar{S}_\psi(t)]^N.
\end{equation}
Note that in the limit $N\to\infty$ the integral in the latter
expression is evidently dominated by the behavior of $\bar{S}_\psi(t)$
and, hence, of $\psi(t)$ in the vicinity of $t=0$.

Two remarks are in order:

(i) One can easily show that $\bar{S}_\psi(t)$ is a positive function that
monotonously decreases from unity at $t=0$ to zero as $t\to\infty$ (see
App.~\ref{sec:auxil}). As a consequence, one can interpret $\bar{S}_\psi(t)$
as the survival probability of an effective particle whose diffusive search
for the target already incorporates the delayed entrance. In other words,
one can introduce an effective FPT $\ttau_k$ via $\bar{S}_\psi(t)=\P\{\ttau_k
>t\}$ that accounts for the extended injection. In this way, the search by
the particles that were injected progressively via the profile $\psi(t)$, is
equivalent to the search by the effective particles injected instantaneously,
i.e., the distribution of the fFPT $\T_N$ is getting close to $\P\{\bar{\T}_N
>t\}=\bar{S}_N(t)$ of the minimum $\bar{\T}_N=\min\{\ttau_1,\ldots,\ttau_N\}$
for large $N$.

(ii) For an instantaneous injection of $N$ particles with a
$\delta$-pulse, one has $\psi(t)\equiv\delta(t)$, so that
$\bar{S}_\psi(t)\equiv S(t)$, i.e., the survival probability of the
target in the presence of just a single searcher. This gives
straightforwardly $\bar{S}_N(t)=[S(t)]^N$, i.e., the standard starting
point of the analyses of the fFPTs in systems with instantaneously
generated $N$ particles
\cite{Weiss83,Lawley20,Lawley20b,Lawley20c,Madrid20,Grebenkov20a,Grebenkov22,LawleyBook}.
We will comment on the results of these analyses in what follows,
comparing them with our theoretical findings for an extended-in-time
injection.

\subsection{Random injection}

In this scenario, we consider the entrance times $\delta_k$ as independent
and identically distributed random variables governed by the PDF $\psi(t)$,
which is determined by a given injection profile. Due to the factorized form
of Eq.~\eqref{z}, the averaging of this expression over the variables $\delta_k$
can be performed straightforwardly, yielding
\begin{equation}
\langle\P\{\T_N>t\}\rangle_{\delta_k}=S_N(t)=[S_\psi(t)]^N,
\end{equation} 
where
\begin{eqnarray}
\nonumber
S_\psi(t)&=&\int^{\infty}_0 dt'\psi(t')S(t - t')\\
\nonumber
&=&\int^{\infty}_0dt'\psi(t')-\int^{\infty}_0dt'\psi(t')[1-S(t-t')]\\
&=&1-\int^t_0 dt'\psi(t')[1-S(t-t')],
\label{Spsi}
\end{eqnarray}
where we took advantage of the fact that $S(t<0)=1$. This relation also implies
\begin{equation}
\label{eq:Hpsi}
H_\psi(t)=\int\limits_0^tdt'\,\psi(t')\,H(t-t'),
\end{equation}
as expected for the sum $\tau_k+\delta_k$ of two independent random variables.
Consequently, in virtue of Eq.~\eqref{HH}, the fFPT-PDF in this scenario obeys
\begin{equation}  
\label{H2}
H_N(t)=-\frac{d}{dt}[S_\psi(t)]^N, 
\end{equation}
while the mean fFPT is determined by the integral
\begin{equation}
\label{mean2}
\langle\T_N\rangle=\int^{\infty}_0dt\,[S_\psi(t)]^N.
\end{equation}
Consequently, in the limit $N\to\infty$ the integral in
Eq.~\eqref{mean2} is also dominated by the behavior of $S_\psi(t)$
and, hence, of $\psi(t)$ in the vicinity of $t=0$. This is the reason
why, although the random injection is formally different from the
deterministic injection, the results of both injection scenarios will
appear very similar in the limit of large $N$, as we proceed to show
in the following.

\subsection{Asymptotic behavior for instantaneous injection}

To set the scene, we first briefly recall the main results corresponding to
the situation in which all particles are simultaneously injected into the
domain and then move independently of each other, searching in parallel for
a given immobile target. Note that the notations and the main equations
presented in Sec.~\ref{sec:1} are valid in this case and correspond to the
choice $\psi(t)=\delta(t)$. To avoid confusion, we add the superscript "0"
when referring to the characteristic properties in this case, such as the
mean fFPT $\T_N^0$, the survival probability $S_N^0(t)$ and the fFPT-PDF
$H_N^{0}(t)$.

In the context of the diffusive motion of particles in a one-dimensional
system, the statistics of the mean fFPT for simultaneous injection of $N$
particles was first addressed by Weiss {\it et al.} \cite{Weiss83}, who
revealed the extremely slow dependence of the mean fFPT on the number of
particles,
\begin{equation}
\label{weiss}
\langle\T_N^0\rangle\simeq\frac{C}{\ln N}\quad(N\to\infty),
\end{equation}
where $C=\ell^2/(4D)$ is the natural time-scale of the search process,
expressed in terms of the diffusion coefficient $D$ and the distance $\ell$
between the starting point and the target. Therefore, the speed up of the
search process due to multiple independent particles starting simultaneously
from the same point turns out to be minor, an observation which provoked
speculations why the numbers of searchers employed in diverse processes in
biology and physiology are typically so high (the so-called "redundancy
principle") \cite{Schuss19}.

Many other aspects of this challenging problem were also discussed in
this seminal paper \cite{Weiss83}. Several extensions and more
elaborate descriptions were presented in subsequent works
\cite{Basnayake19,Lawley20,Lawley20b,Lawley20c,Madrid20,Grebenkov20a},
including, e.g., the analysis of the statistics of the $k$th fastest
FPT and its higher-order and joint moments, as well as a
generalisation of the asymptotic result \eqref{weiss} for diffusion in
higher-dimensional spaces. Moreover, it was demonstrated that in the
large $N$ limit, the fFPT-PDF converges to the universal Gumbel
distribution, with non-universal, dynamic-specific scale and location
parameters $a_N$ and $b_N$ \cite{Lawley20}. The large-$N$ asymptotic
behavior of these parameters, as well as that of the fFPT moments,
appears to be quite elaborate and completely dominated by the
short-time behavior of the survival probability $S(t)$. In particular,
Lawley derived the asymptotic expressions for $a_N$ and $b_N$ and thus
for the mean and the variance of the fFPT under the rather general
assumption on the short-time asymptotic behavior of the survival
probability,
\begin{equation}
\label{eq:St0_cond}
1-S(t)\sim A\,t^\alpha \,e^{-C/t}\quad(t\to 0),
\end{equation}
with the constants $A>0$ and $C>0$, and $\alpha\in\R$ \cite{Lawley20}. We note
parenthetically that the expression for the mean fFPT in Eq.~\eqref{weiss}
stems exactly from the exponentially fast decay of $1-S(t)$ above, while the
precise value of the exponent $\alpha$ in the pre-exponential factor is less
important. In turn, it was shown that if the initial particle distribution is
uniform, the mean fFPT scales as $1/N^2$ for perfect reactions upon the first
arrival to the target, and as $1/N$ for the case of finite reactivity
\cite{Madrid20,Grebenkov20a}, entailing a much more significant speed up of
the search process---as compared to the $1/\ln N$ law, which occurs when all
particles start from the same location. We finally note that interactions
between particles or correlations between their first-passage times may
totally change the statistics of the fFPT (see \cite{Bray13,Agranov18} and
references therein).

\subsection{Asymptotic behavior for extended injection}
\label{sec:3}

Returning to the case of an extended-in-time injection, we focus on the limit
$N\to\infty$, for which the behavior of the mean fFPT is entirely dominated
by the short-time behavior of the survival probabilities $\bar{S}_\psi(t)$ and
$S_\psi(t)$ and, hence, by the form of the injection profile $\psi(t)$ in the
vicinity of $t=0$. Assuming that $\psi(t)$ originates from a random transport
processes, e.g., the random motion of particles within a nuclear pore or an
ion channel, we stipulate here that it has the quite generic form
\begin{equation}
\label{eq:psi_short}
\psi(t)\approx a\, t^{\nu-1}\, e^{-(c/t)^\mu}\quad(t\to 0),
\end{equation} 
with strictly positive constants $a$ and $c$ and exponents $\nu$ and $\mu\geq0$.
In App.~\ref{sec:short}, we show the equivalence of $\bar{S}_\psi(t)$ and $S_
\psi(t)$ to leading order for the large-$N$ limit, such that it suffices to
consider only either of these quantities. This implies that the mean fFPTs in
both scenarios, defined in Eqs.~\eqref{mean1bar} and \eqref{mean2}, are equal
to each other to leading order: $\langle\bar{\T}_N\rangle=\langle\T_N\rangle$,
as $N\to\infty$. Moreover, we find the following asymptotic behavior (see
App.~\ref{sec:short} for more details)
\begin{equation}  \label{eq:St_cond}
1-S_\psi(t)\approx\bar{A} \,t^{\bar{\alpha}}\,e^{-(\bar{C}/t)^{\bar{\mu}}}\quad
(t\to0),
\end{equation}
where the coefficients $\bar{A}$, $\bar{C}$, $\bar{\alpha}$, and
$\bar{\mu}$ are determined by the parameters in
Eqs.~(\ref{eq:St0_cond}) and (\ref{eq:psi_short}), as will be
discussed below. Note that the crucial parameter
$\bar{\mu}$ is given by the value of the exponent that characterizes
the more singular behavior of the search dynamics: either the
time-extended injection process $\psi(t)\sim\exp(-1/t^{\mu})$, or the
diffusive transport of particles to the target, $1
-S(t)\sim\exp(-1/t^\eta)$ (with $\eta=1$); in fact, it controls the
overall behavior of the survival probability $S_\psi(t)$ in the limit
$t\to0$. We also note parenthetically that $\eta=1$ is a fingerprint
feature of Brownian motion and, in fact, $\eta$ may differ from unity
for anomalous diffusion dynamics. In particular, for any Gaussian
process the exponent $\eta$ characterizing the singularity in the
short-$t$ asymptotic of the FPT-PDF and, hence, of the survival
probability, coincides with the so-called anomalous diffusion exponent
\cite{baruch}. Clearly, in the challenging and sometimes more
physically-realistic case of anomalous diffusion, which takes place at
least at sufficiently short times, one will encounter a much richer
scenario than in the case of a Brownian motion. Below we restrict our
analysis to the case of Brownian motion, showing that even in this
simpler case three different scenarios may be distinguished.

The asymptotic behavior (\ref{eq:St_cond}) determines the large-$N$ limit of
the mean fFPT and related quantities. We sketch here simple arguments for the
mean fFPT $\langle\T_N\rangle$ that can be made more rigorous and further
extended by the asymptotic tools discussed in \cite{Weiss83,Lawley20}. The
function $S_N(t)=[S_\psi(t)]^N$ is a monotonically decreasing function of
$t$ such that the integral in Eq.~\eqref{mean2} is dominated by the region
in the vicinity of $t=0$. This function is also a monotonically decreasing
function of $N$, which implies that the region around $t=0$ shrinks upon
increase of $N$. In a first approximation, one can evaluate the integral in
Eq.~\eqref{mean2} by replacing $[S_\psi(t)]^N$ by a Heaviside step function,
$\Theta(T-t)$, which is equal to unity for $t<T$ and zero otherwise. The mean
fFPT is thus approximately equal to the cutoff value $T$ delimiting the range
$(0,T)$ of the contributing times which can be set by the condition $[S_\psi
(T)]^N=1-q_0$, where $q_0$ is an auxiliary parameter determining a "sufficient"
drop of $[S_\psi(t)]^N$. Substituting into Eq.~(\ref{eq:St_cond}), one finds
$\bar{A}T^{\bar{\alpha}}\exp(-(\bar{C}/T)^{\bar{\mu}})\approx1-(1-q_0)^{1/N}
\approx q/N$, for sufficiently large $N$; here, $q=-\ln(1-q_0)$. An approximate
solution of this transcendental equation can be found perturbatively by taking
the logarithm such that
\begin{eqnarray*}
\langle\T_N\rangle&\approx&T\approx\frac{\bar{C}}{\bigl[\ln(N)+\alpha\ln(T)+
\ln(\bar{A}/q)\bigr]^{1/\bar{\mu}}}\\
&\approx&\frac{\bar{C}}{(\ln N)^{1/\bar{\mu}}}\biggl(1+\frac{\frac{\bar{
\alpha}}{\bar{\mu}}\ln\ln N-\ln(\bar{A}\bar{C}^{\bar{\alpha}}/q)}{\bar{\mu}
\ln N}\biggr),
\end{eqnarray*} 
where $\ln(T)$ in the denominator in the top equation was replaced by its
leading-order expression $\ln(\bar{C}/(\ln N)^{1/\bar{\mu}})$. Importantly,
the leading-order term does not depend on the somewhat arbitrary parameter
$q$. For $\bar{\mu}=1$, we retrieve the expression for the mean fFPT derived
by Lawley for the instantaneous injection scenario via a considerably more
elaborate and rigorous asymptotic analysis \cite{Lawley20} (see also
App.~\ref{sec:Lawley}). Moreover, if we choose $q=e^{-\gamma}\approx0.5615$,
where $\gamma$ is the Euler constant, the correction term becomes identical
to that of Lawley's rigorous expression, thus confirming that our rough
approximation captures correctly the subtle behavior of the mean fFPT. In
other words, our expansion
\begin{equation}
\label{eq:meanFPT_asympt0}
\langle\T_N\rangle\approx\frac{\bar{C}}{(\ln N)^{1/\bar{\mu}}}\biggl(1+
\frac{\frac{\bar{\alpha}}{\bar{\mu}}\ln\ln N-\ln(\bar{A}\bar{C}^{\bar{
\alpha}}e^{\gamma})}{\bar{\mu}\ln N}\biggr) 
\end{equation}
is reduced to the rigorous result by Lawley in the special case of an
instantaneous injection.

The parameters $\bar{A}$, $\bar{C}$, $\bar{\alpha}$, and $\bar{\mu}$ in
Eq.~(\ref{eq:St_cond}) can be expressed in terms of the parameters $A$, $C$,
and $\alpha$ of the survival probability $S(t)$, and $a$, $c$, $\nu$, and
$\mu$ of the injection profile $\psi(t)$. Skipping technical details (see
App.~\ref{sec:short}), we draw the main conclusions for three different types
of the injection profile:

\textbf{(i)} If $0\leq\mu<1$, the survival probability $S(t)$ is the limiting
factor such that 
\begin{equation}
\label{eq:parameters1}
\bar{\mu}=1,\quad\bar{C}=C,   
\end{equation}
and the leading term $C/\ln N$ does not depend on the shape of the injection
profile. The latter only affects the correction term through the coefficients
$\bar{A}$ and $\bar{\alpha}$ given by Eqs.~(\ref{eq:coeff_mu0}) for $\mu=0$
and by Eqs.~(\ref{eq:coeff_mu0_1}) for $0<\mu<1$.

\textbf{(ii)} If $\mu=1$, relations (\ref{eq:St0_cond}) and  (\ref{eq:psi_short})
exhibit similar asymptotic behaviors, yielding $\bar{\mu}=1$ and $\bar{C}=(
\sqrt{C}+\sqrt{c})^2$ for the coefficient in front of the leading term (the
correction term also changes, see Eqs.~(\ref{eq:Abar_case2}) for $\bar{A}$
and $\bar{\alpha}$).

\textbf{(iii)} If $\mu>1$, the profile $\psi(t)$ decreases faster than $1-S(t)$
at short times such that the extended injection is the limiting factor for the
fastest arrival to the target. In this case, the leading term of $\langle\T_N
\rangle$ exhibits the even slower decrease of the form $\bar{C}/(\ln N)^{1/\bar{
\mu}}$ with $\bar{C}=c$ and $\bar{\mu}=\mu$, which is independent of the
diffusive dynamics, that only affects the correction term (see
Eqs.~(\ref{eq:Abar_case1}) for $\bar{A}$ and $\bar{\alpha}$).

In a similar way, one can analyze the higher-order moments of the fFPT-PDF for
an extended injection. We restrict our attention to the case $0\leq\mu\leq1$,
for which $\bar{\mu}=1$ and thus the short-time asymptotic behavior of the
survival probability $S_\psi(t)$ admits the same form as Eq.~(\ref{eq:St0_cond}),
upon replacing $C$, $\alpha$, and $A$ by $\bar{C}$, $\bar{\alpha}$, and $\bar{
A}$, respectively. As a consequence, one can apply the original result of Lawley
for the instantaneous injection case (see App.~\ref{sec:Lawley}). In particular,
if $\bar{\alpha}>0$, the variance of the fFPT is given by
\begin{equation}
\label{eq:TN_var}
\mathrm{Var}\{\T_N\}\approx\frac{\pi^2}{6}\biggl(\frac{\bar{C}}{\bar{\alpha}^2
\bar{W}(1+\bar{W})}\biggr)^2\qquad(N\gg1),
\end{equation}
where 
\begin{equation}
\bar{W}=W_0\bigl(\bar{C}(\bar{A}N)^{1/\bar{\alpha}}/\bar{\alpha}\bigr),
\end{equation}
and $W_0(z)$ is the principal branch of the Lambert W-function \cite{Lawley20}
(see App.~\ref{sec:Lawley} for other cases). For sufficiently  large $N$, one
can use the asymptotic behavior of this function to get
\begin{equation}
\label{eq:TN_var0}
\mathrm{Var}\{\T_N\}\approx V_N,  \qquad V_N = \frac{\pi^2}{6}\,\frac{C^2}{(\ln N)^4}\qquad(N\gg1),
\end{equation}
i.e., the leading-order behavior of the variance is independent of the
injection profile.

\section{Numerical analysis of a particular example}
\label{sec:4}

To illustrate the above-discussed effects and to check the accuracy of the
derived asymptotic relations, we consider diffusion in the upper half-space
towards an absorbing plane (the target). As lateral displacements do not
affect the transverse motion and thus the statistics of the first arrival
onto the plane, this first-passage problem is equivalent to diffusion on the
half-line $\Omega=(0,+\infty)$ from a fixed starting point $x_0>0$ towards an
absorbing origin. In this prototypical setting, one has $S(t)=\erf(x_0/\sqrt{
4Dt})$, where $\erf(z)$ is the error function, so that the asymptotic relation
(\ref{eq:St0_cond}) holds, with
\begin{equation}
C=\frac{x_0^2}{4D},\qquad\alpha=\frac{1}{2},\qquad A=\frac{1}{\sqrt{\pi C}}.
\end{equation}
Despite the simplicity of this example, the underlying L\'evy-Smirnov
PDF $H(t)=-\tfrac{d}{dt}S(t)=x_0e^{-x_0^2/(4Dt)}/\sqrt{4\pi Dt^3}$
captures well the short-time behavior of the FPT-PDF in various
geometric settings, as well as in higher dimensions, and is thus a
representative example for the large-$N$ asymptotic behavior of the
fFPT statistics.

\begin{figure}
\begin{center}
\includegraphics[width=85mm]{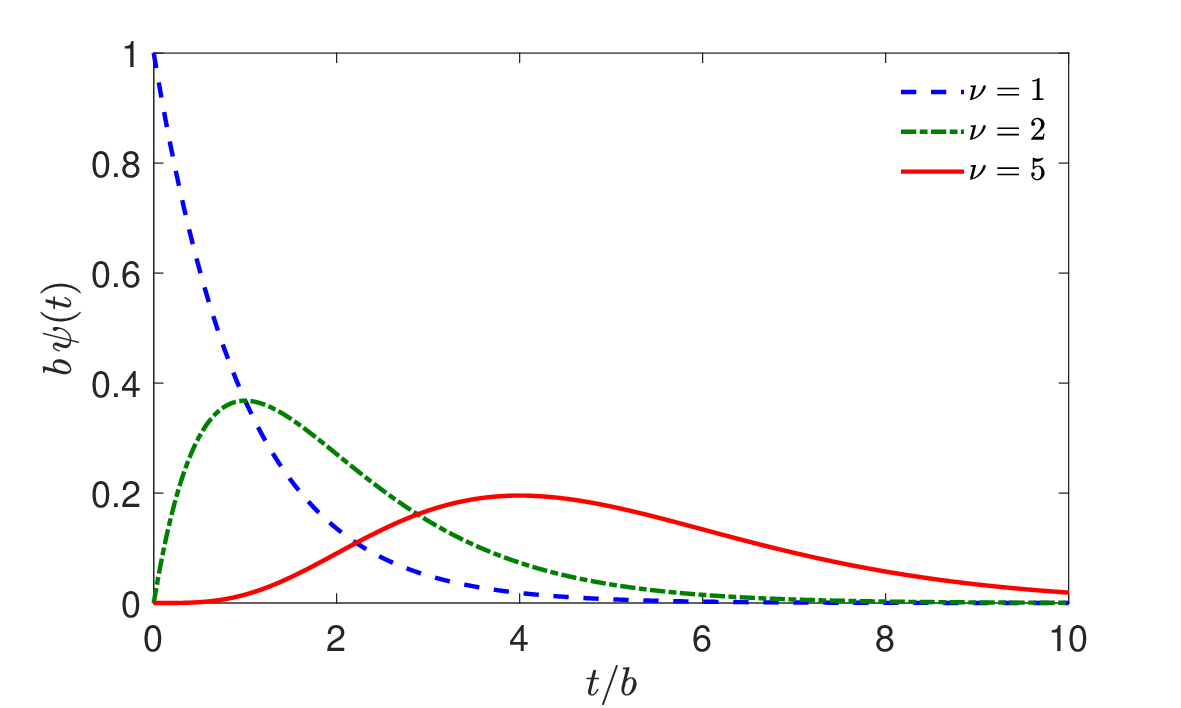}  
\end{center}
\caption{
Three injection profiles modeled by the Gamma distribution
(\ref{eq:gGamma}), with the timescale $b=1$ and three scale parameters
$\nu$ indicated as in the legend.}
\label{fig:profile_gamma}
\end{figure}

In turn, the entrance times $\delta_k$ are modeled via the gamma distribution
that corresponds to the injection profile
\begin{equation}
\label{eq:gGamma}
\psi(t)=\frac{t^{\nu-1}e^{-t/b}}{b^\nu\Gamma(\nu)},
\end{equation}
with the shape parameter $\nu>0$ and the scale parameter $b>0$. This
model obeys the short-time behavior (\ref{eq:psi_short}) with $\mu=0$
and $a=b^{ -\nu}/\Gamma(\nu)$ so that this injection profile is
supposed to have the {\it weakest} impact onto the fFPT (case
\textbf{(i)}); indeed, according to Eq.~(\ref{eq:parameters1}), the
leading-order behavior of the mean fFPT remains unchanged. We have
chosen this model to illustrate that even such a mild modification in
the short-time asymptotic behavior of the survival probability
$S_\psi(t)$ can produce significant alterations in the behavior of the
fFPT.  Figure \ref{fig:profile_gamma} illustrates three injection
profiles modeled by Eq.~(\ref{eq:gGamma}). Note that $\psi(t)$
formally converges to $\delta (t)$ in the limit $\nu\to0$,
corresponding to instantaneous injection. In the following, we mostly
focus on the case $b=1$ and $\nu=5$ as a representative example.
In App. \ref{sec:uniform}, we discuss another common injection
profile when the delay times $\delta_k$ are uniformly distributed over
an interval $(0,T)$. In this setting, we get exact explicit formulas
for $H_\psi(t)$ and $S_\psi(t)$ and reveal the role of the shape of
the injection profile.

To check the asymptotic relations, we need to compute numerically the
mean fFPT $\langle\T_N\rangle$ and the PDF $H_N(t)$. For this purpose,
we introduce a time cutoff $t_{\rm max}$ to be tenfold larger than
either of the timescales $C=x_0^2/(4D)$ and $b$, discretize the time
interval $[0,t_{\rm max}]$ with a small time step $\delta$, and
compute $S_\psi(t)$ at times $n\delta$ (with $n = 1,2,\ldots,t_{\rm
max}/\delta$) via a fast numerical convolution of the integral in
Eq.~(\ref{Spsi}). From the knowledge of $S_\psi(t)$, one can easily
access both the PDF $H_N(t)$ and the mean fFPT $\langle
\T_N\rangle$ via numerical integration in Eq.~(\ref{mean2}).
Similarly, we evaluate numerically $\bar{S}_\psi(t)$ from
Eq.~(\ref{Spsibar}) and thus access $\bar{H}_N(t)$ and
$\langle\bar{\T}_N\rangle$. Varying both $t_{\rm max}$ and $\delta$,
we can control the accuracy of this numerical computation. In the
following, we refer to the obtained PDF and the mean as "exact"
results that will be compared to explicit but approximate asymptotic
relations.

\begin{figure}
\begin{center}
\includegraphics[width=85mm]{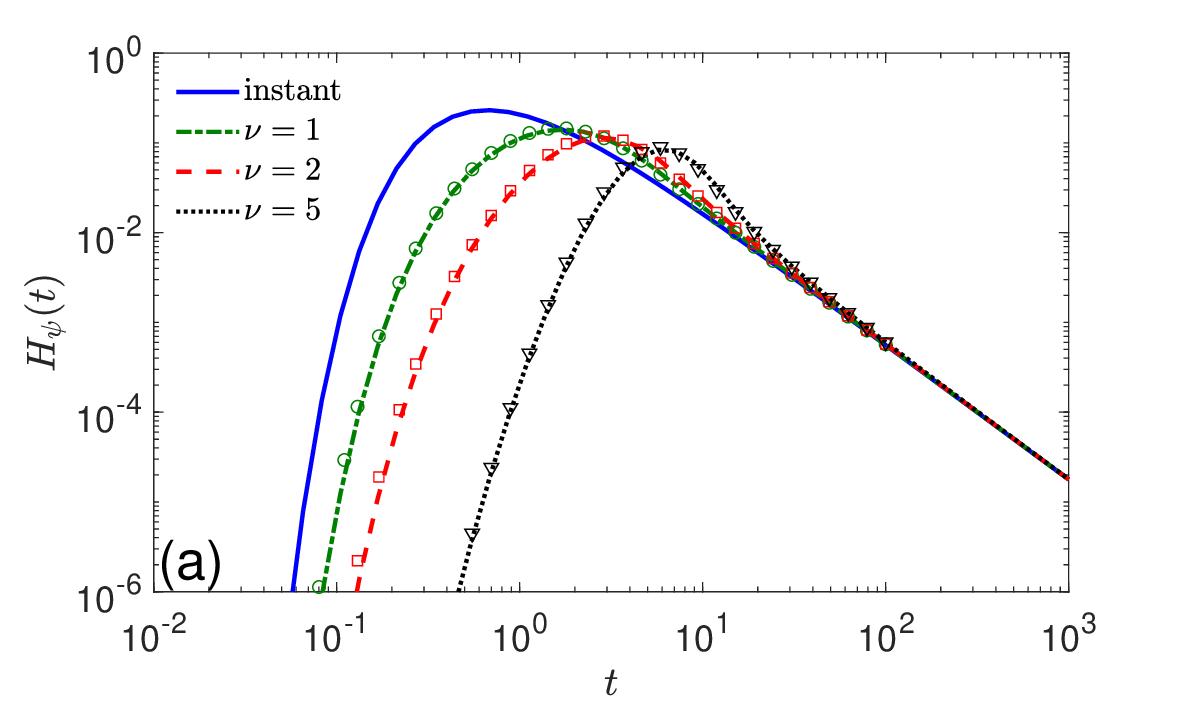} 
\includegraphics[width=85mm]{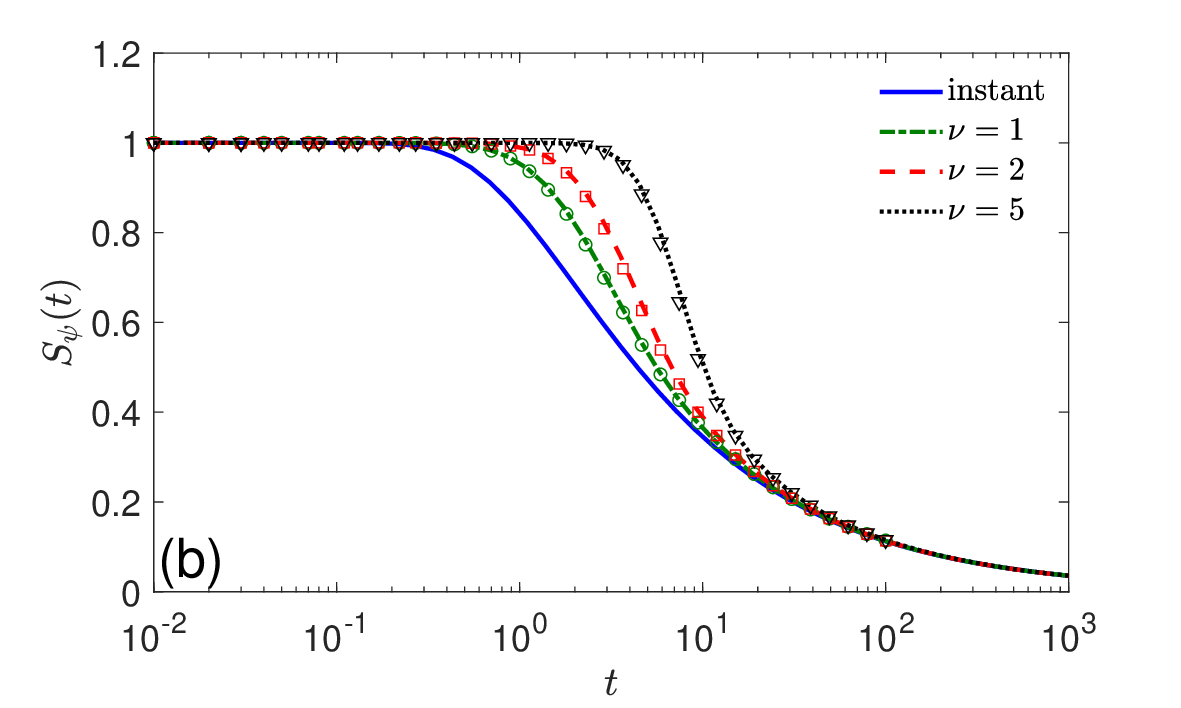} 
\end{center}
\caption{{\bf (a)} PDF $H_\psi(t)$ and {\bf (b)} the survival probability
$S_\psi(t)$ of the delayed FPT $\tau_k+\delta_k$ for diffusion on the
half-line, with $x_0=2$, $D=1$, and the injection profile given by the
gamma model (\ref{eq:gGamma}), with $b=1$ and three values of
$\nu$. The solid line represents $H(t)$ and $S(t)$ of an instantaneous
injection ($\nu =0$). The symbols show $\bar{H}_\psi(t)$ and
$\bar{S}_\psi(t)$ for the deterministic injection with the same set of
parameters.}
\label{fig:Hprime}
\end{figure}

Let us first inspect the effect of an extended injection modeled by the gamma
distribution (\ref{eq:gGamma}), with $b=1$ and three different values of $\nu$.
Figure \ref{fig:Hprime}(a) illustrates the PDF $H_\psi(t)$ of the delayed FPT
$\tau_k+\delta_k$ for a single particle. As the profiles with larger $\nu$
delay the entrance, the corresponding PDFs are shifted to longer times. In
turn, the limit $\nu\to0$ formally corresponds to the instantaneous injection
characterized by $H(t)$. One sees therefore how the extended injection modifies
$H(t)$. The mean value is infinite, $\langle\tau_k+\delta_k\rangle=\infty$,
while the most probable value grows with $\nu$ (note that $\langle\delta_k
\rangle=\nu b$ and $T_{\rm mp}=(\nu-1)b$). For comparison, we also show the
PDF $\bar{H}_\psi(t)$ of the effective FPT $\bar{\tau}_k$. One sees that it
is almost identical with $H_\psi(t)$ for the considered range of parameters
and times. Figure \ref{fig:Hprime}(b) presents the corresponding survival
probabilities $S_\psi(t)$ and $\bar{S}_\psi(t)$ for a single particle.

\begin{figure}
\begin{center}
\includegraphics[width=85mm]{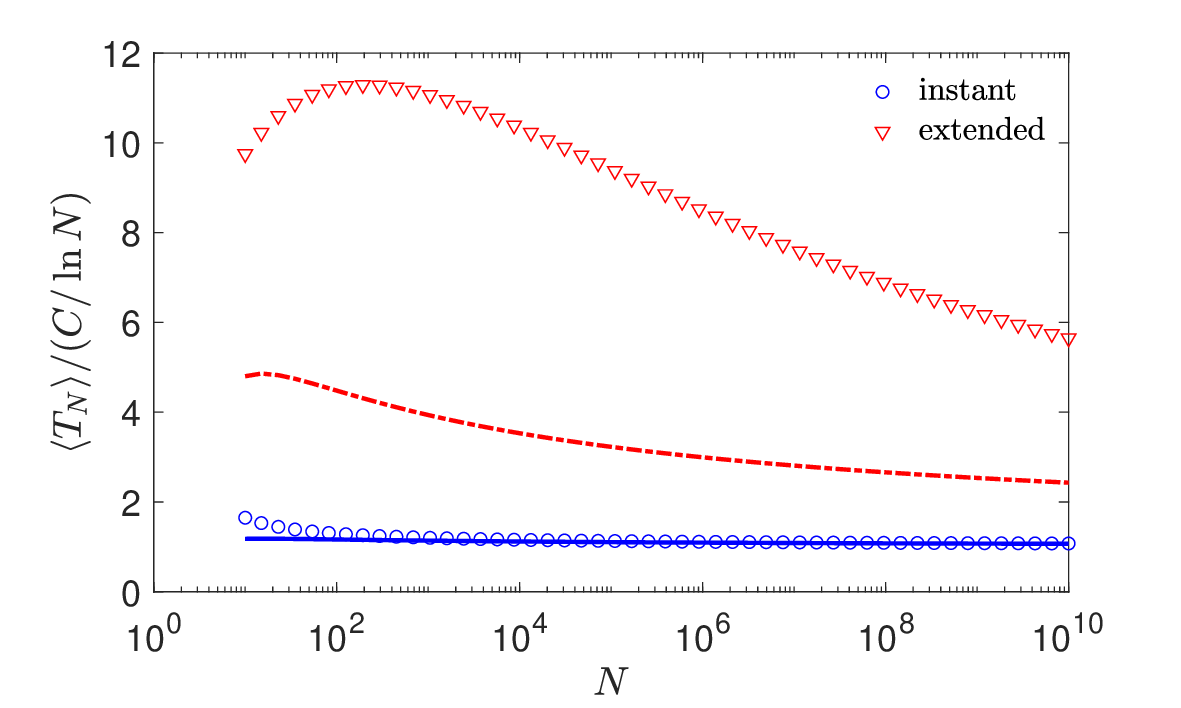} 
\end{center}
\caption{Mean fFPTs $\langle\T_N^0\rangle$ and $\langle\T_N\rangle$, rescaled
by $C/\ln N$, as functions of $N$ for diffusion on the half-line, with $x_0=2$,
$D=1$ (and thus $C=1$ and $A=1/\sqrt{\pi}$), for an instantaneous injection
(circles) and an extended injection of $N$ particles modeled by the gamma
distribution (\ref{eq:gGamma}) with $b=1$ and $\nu=5$ (triangles). The symbols
show the mean fFPTs obtained by computing numerically the convolution in
Eq.~(\ref{Spsi}) and then evaluating the integral in Eq.~(\ref{mean2}), used
as benchmarks. The lines present the asymptotic relation
(\ref{eq:meanFPT_asympt0}), with $\bar{\mu}=1$, $\bar{C}=1$, and two settings:
$\bar{\alpha}=\alpha+2\nu=10.5$ and $\bar{A}=A(Cb)^{-\nu}$ for the extended
injection (dashed line), and $\bar{\alpha}=\alpha=0.5$ and $\bar{A}=A$ for
the instantaneous injection (solid line). Overall, a considerable increase of
the mean fFPT, by an order of magnitude, is observed for the case of the
time-extended injection in this setting.}
\label{fig:Tmean_1d_gamma}
\end{figure}

Figure \ref{fig:Tmean_1d_gamma} compares the mean fFPTs $\langle\T_N^0\rangle$
and $\langle\T_N\rangle$ for instantaneous and extended injection of $N$
particles modeled by the gamma distribution with $b=1$ and $\nu=5$. For the
case of an instantaneous injection, the mean fFPT $\langle\T_N^0\rangle$ is
accurately approximated by the asymptotic relation (\ref{eq:meanFPT_asympt0}),
even for $N$ as small as $100$; moreover, the latter is close to the leading
term $C/\ln N$. The good quality of this asymptotic approximation for
diffusion on the half-line was reported earlier in \cite{Lawley20}. In
contrast, the quality of this approximation is considerably worse for an
extended injection with $\nu=5$. On the one hand, we showed above that the
survival probability $S_\psi(t)$ satisfies the asymptotic behavior
(\ref{eq:St_cond}) with $\bar{\mu}=1$, allowing one to apply Lawley's
asymptotic results. In particular, the asymptotic relation
(\ref{eq:meanFPT_asympt0}) holds with the unchanged leading order $C/\ln N$
and the correction term with modified parameters $\bar{\alpha}=\alpha+2\nu$
and $\bar{A}=A(Cb)^{-\nu}$. One sees that this asymptotic relation indeed
approaches the exact mean fFPT in the limit $N\to\infty$. On the other hand,
the approach to this limiting behavior is extremely slow, and very large $N$
are needed to get an accurate approximation (e.g., deviations are still
noticeable even at $N=10^{10}$).

\begin{figure}
\begin{center}
\includegraphics[width=85mm]{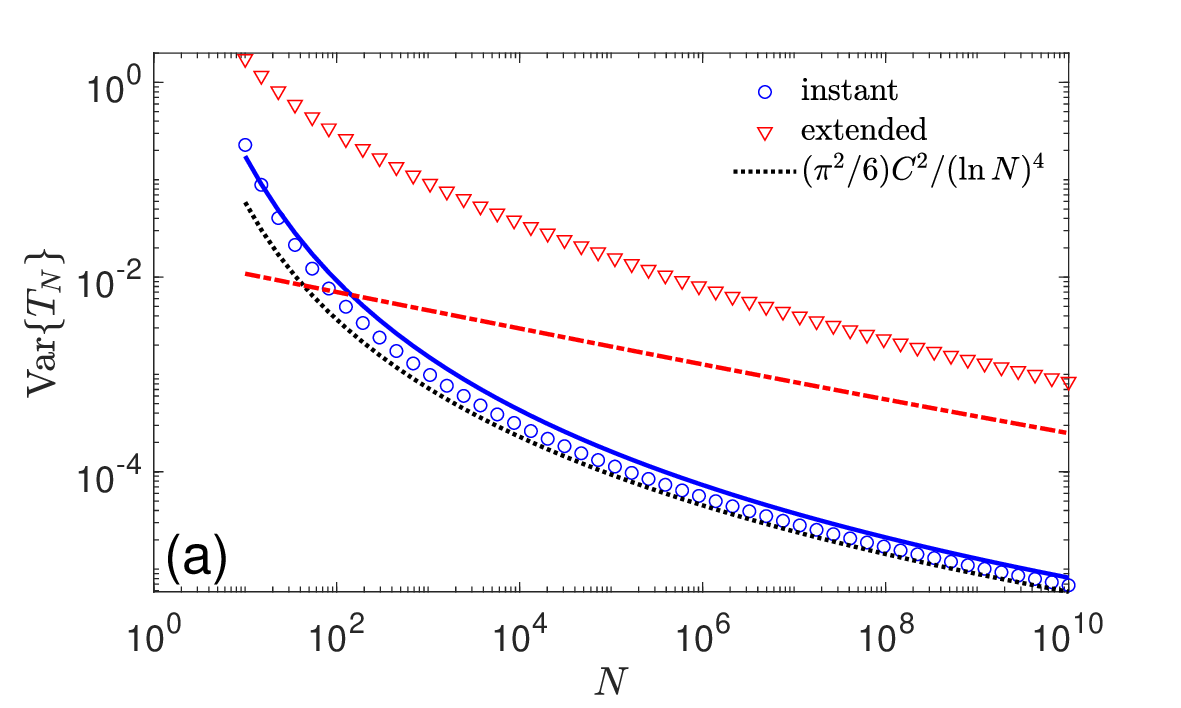} 
\includegraphics[width=85mm]{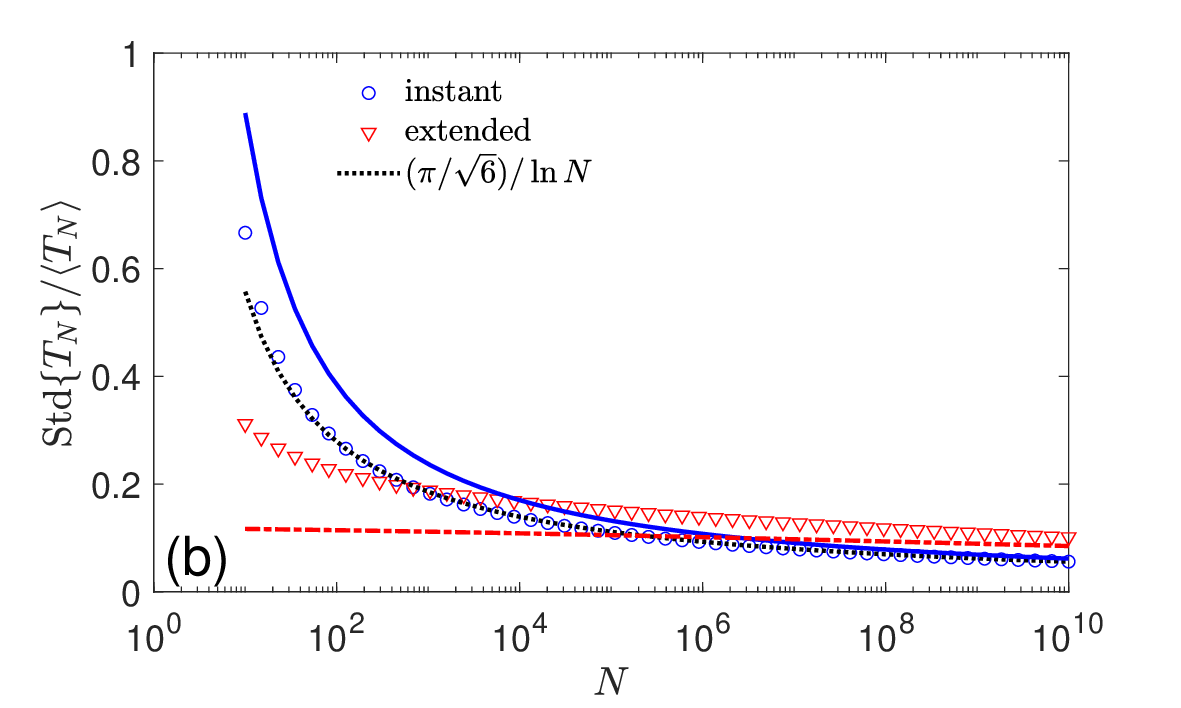} 
\includegraphics[width=85mm]{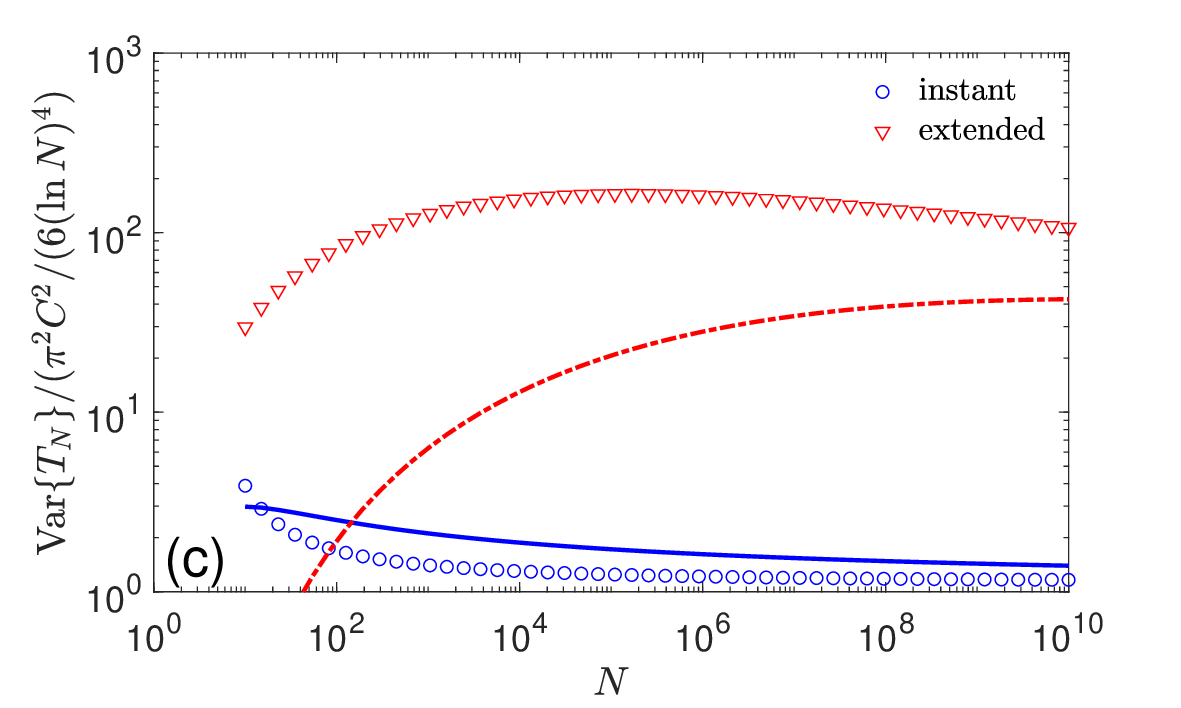} 
\end{center}
\caption{{\bf (a)} Variance of the fFPT $\T_N$ as a function of $N$ for
diffusion on the half-line, with $x_0=2$, $D=1$ (and thus $C=1$ and $A=1/
\sqrt{\pi}$), with an extended injection of $N$ particles modeled by the
gamma distribution (\ref{eq:gGamma}) with $b=1$ and $\nu=5$. For
comparison, the variance of the fFPT $\T_N^0$ for an instantaneous
injection is shown.  The symbols show the exact variance, obtained
from computing numerically the convolution in Eq.~(\ref{Spsi}) and
then by evaluating the first and second moments of $\T_N$ via the
integrals $\int\nolimits_0^\infty dt\, [S_\psi(t)]^N$ and
$2\int\nolimits_0^\infty dt\,t[S_\psi(t)]^N$. The solid and dashed
lines present Lawley's asymptotic relation (\ref{eq:TN0_var}), in
which $a_N$ is given by Eq.~(\ref{eq:aNbN_Lawley}), either with
parameters $C$, $\alpha$, and $A$ for the instantaneous injection
(solid line), or with parameters $\bar{C}=C$,
$\bar{\alpha}=\alpha+2\nu$, and $\bar{A}=A(Cb)^{-\nu}$ for extended
injection (dashed line). Note that the black dotted line indicates the
lowest-order behavior $V_N=\tfrac{\pi^2}{6} C^2/(\ln N)^4$.  {\bf (b)}
The coefficient of variation, $\sqrt{\Var\{\T_N\}}/\langle
\T_N\rangle$, for the same two settings of instantaneous and extended
injections, with the black dotted line indicating the lowest-order
behavior.  {\bf (c)} The variance rescaled by its leading-order term
$V_N$ given by Eq. (\ref{eq:TN_var0}).}
\label{fig:Tmean_1d_gamma2}
\end{figure}

Figure \ref{fig:Tmean_1d_gamma2}(a) presents the variance of the fFPT
as a function of the overall number $N$ of injected particles for the
cases of instantaneous and extended-in-time injections. We observe
that for an extended injection the variance is nearly two orders of
magnitude larger than its counterpart for the instantaneous injection,
meaning that the fFPT exhibits much stronger fluctuations in the
former case than in the latter one.  This emphasizes that, apparently,
the knowledge of the mean fFPT alone is insufficient to gain a full
understanding of the behavior in such a system and the analysis of the
PDF is highly desirable.  The panel (b) presents the coefficient of
variation, defined as the ratio of the standard deviation of $\T_N$
and its mean value, $\sqrt{\Var\{\T_N\}}/\langle \T_N\rangle$.  For
the case of an extended injection, the coefficient of variation is
somewhat larger than for the case of an instantaneous injection, but
this difference is quite modest.

The last panel of Fig.~\ref{fig:Tmean_1d_gamma2} presents the
variance, rescaled by its limiting behavior
$V_N=\tfrac{\pi^2}{6}C^2/(\ln N)^4$ at very large $N$. For the case of
instantaneous injection, this ratio is close to unity for both the
exact variance and its asymptotic form (\ref{eq:TN_var}), as
expected. In turn, for the extended injection, the ratio
$\Var\{\T_N\}/V_N$ exceeds unity significantly, by a factor of
hundred, whereas the asymptotic relation (\ref{eq:TN_var}) does not
appear to work either. What is going wrong here? To answer this
question, we recall that Eq.~(\ref{eq:TN_var}) relies on the large-$x$
behavior of the Lambert function, $W_0(x)\approx\ln x-\ln(\ln x)+o(1)$
as $x\to\infty$. In this example, the argument of this function is
$x=(N/\sqrt{\pi})^{1/10.5}/10.5 \leq0.81$ for the whole considered
range of $N$ up to $10^{10}$. In other words, even though the
injection profile does not affect the leading-order behavior (i.e.,
$\bar{C}=C$), the change of $\alpha$ to $\bar{\alpha}=
\alpha+2\nu$ considerably reduces the range of applicability of the above
asymptotic formulas. Moreover, the asymptotic formula (\ref{eq:TN_var}),
which represents only the leading term, is not accurate here because the
subleading terms may provide significant contributions for the considered
range of $N$. Even though the leading term becomes dominant in the limit
$N\to\infty$, it is not sufficient for estimating the variance at
intermediate values of $N$.

\begin{figure*}
\includegraphics[width=85mm]{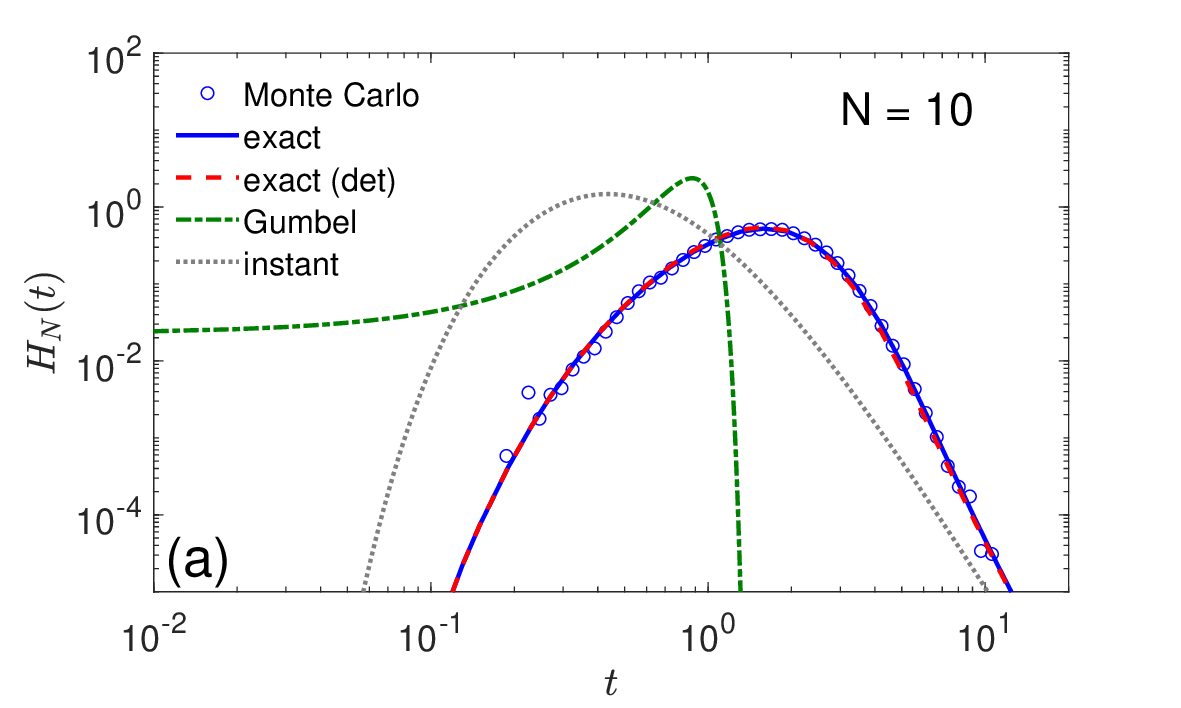} 
\includegraphics[width=85mm]{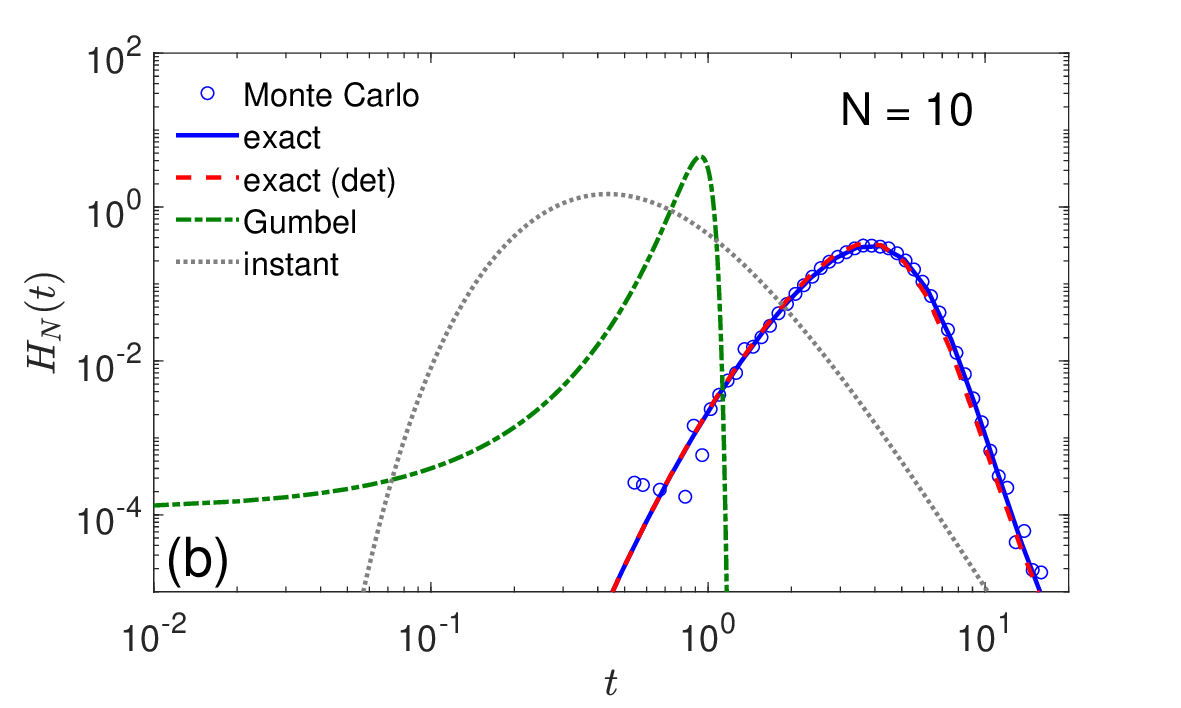} 
\includegraphics[width=85mm]{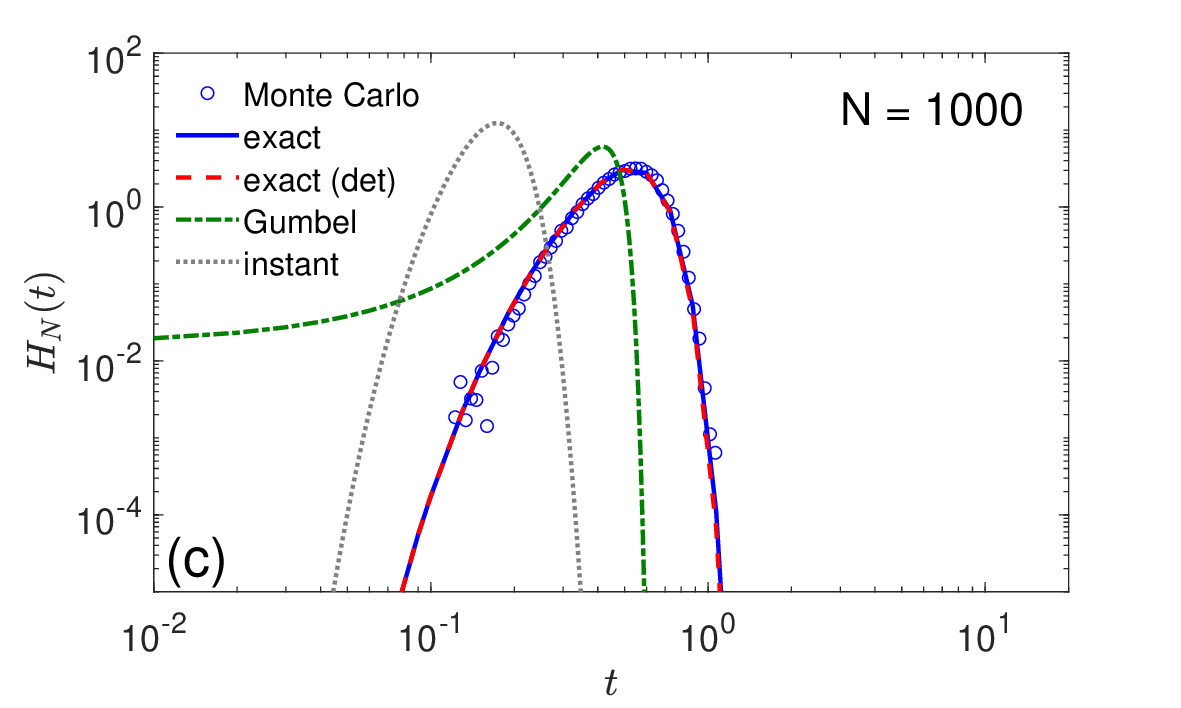} 
\includegraphics[width=85mm]{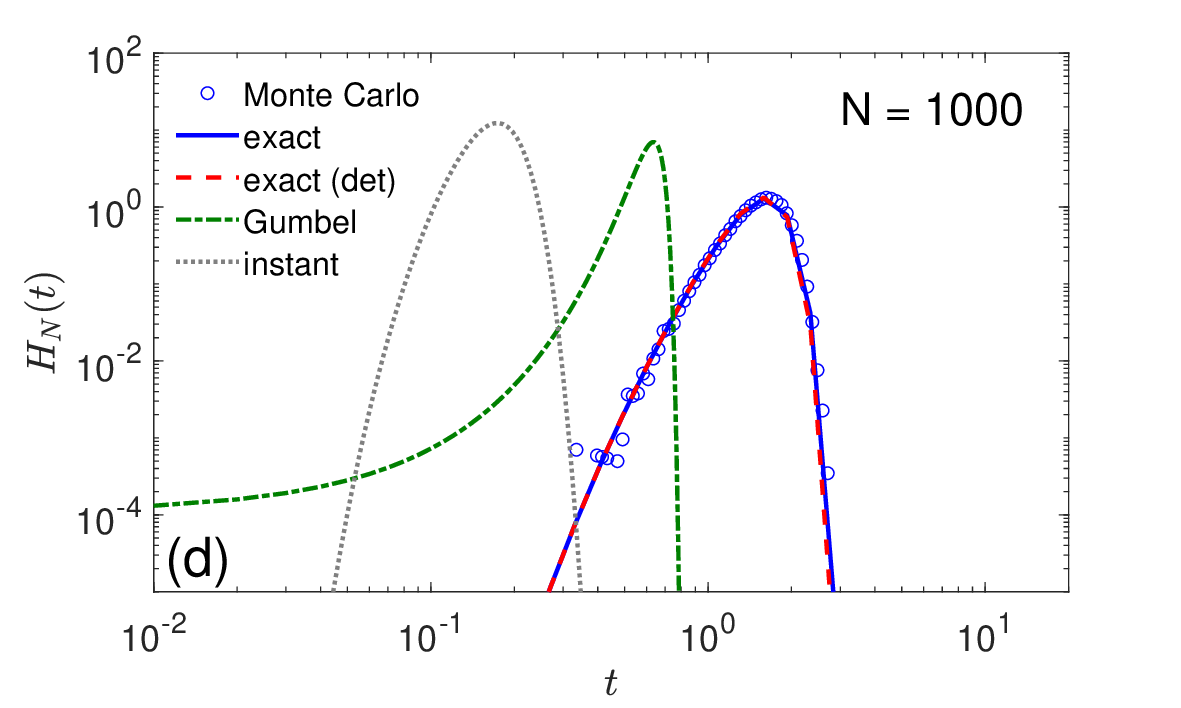} 
\caption{PDF $H_N(t)$ of the fastest FPT $\T_N$ for diffusion on the
half-line, with $x_0=2$, $D=1$ (and thus $C=1$), and an extended
injection modeled by the gamma distribution (\ref{eq:gGamma}) with
$b=1$, and four combinations of $\nu$ and $N$: {\bf (a)} $\nu=2$,
$N=10$; {\bf (b)} $\nu=5$, $N=10$; {\bf (c)} $\nu=2$, $N=1000$, and
{\bf (d)} $\nu=5$, $N= 1000$. The circles represent rescaled
histograms obtained by generating $\T_N$ (with $10^5$
realizations). The solid line represents the exact solution
$H_N(t)=NH_{\psi}(t)[S_{\psi}(t)]^{N-1}$, where $H_{\psi}(t)$ and
$S_{\psi}(t)$ are obtained via fast numerical convolutions. The dashed
line shows the exact solution
$\bar{H}_N(t)=N\bar{H}_{\psi}(t)[\bar{S}_
\psi(t)]^{N-1}$, where $\bar{H}_{\psi}(t)$ and $\bar{S}_{\psi}(t)$ are
obtained via fast numerical convolutions. The dash-dotted line
represents the asymptotic Gumbel distribution, with the scale and
location parameters $a_N$ and $b_N$ given by
Eq.~(\ref{eq:aNbN_Lawley}) with the modified parameters
$\bar{A}=aAC^{-\nu}\Gamma(\nu)$ and $\bar{\alpha}=\alpha+2\nu$, see
App.~\ref{sec:short}. The dotted line indicates the exact distribution
for an instantaneous injection.}
\label{fig:H_1d_gamma_nu5}
\end{figure*}

Figure \ref{fig:H_1d_gamma_nu5} illustrates the PDF $H_N(t)$ of the fastest
FPT $\T_N$ for an extended random injection modeled by the gamma
distribution (\ref{eq:gGamma}) with $b=1$ and two values of the scale
parameter: $\nu=2$ and $\nu=5$. In agreement with the above analysis, the
PDF is shifted to the left (to shorter times) and narrows with increasing $N$.
The exact solution determined from $H_\psi(t)$ and $S_\psi(t)$ shows
excellent agreement with Monte Carlo simulations. Moreover, the PDF $\bar{H}_
N(t)$ of $\bar{\T}_N$ for the deterministic injection with the same profile
is almost indistinguishable from $H_N(t)$, demonstrating the asymptotic
equivalence between $\T_N$ and $\bar{\T}_N$ even for moderate $N$. This
figure highlights two important conclusions:

\textbf{(i)} First, an extended injection of particles substantially delays
the arrival of the fastest particle; in particular, the typical time of the
first arrival that characterizes the maximum of the PDF, is shifted to
longer times as compared to the instantaneous injection of the same number
of particles. For instance, in the case $\nu=5$ and the considered values of
$N$, a tenfold delay is observed. This delay is weaker for $\nu=2$ but it
can be much more substantial for larger $\nu$, i.e., for a faster decay of
$\psi(t)$. As a consequence, neglecting an extended injection, as done in
former works on this topic, may lead to strong misrepresentations. 

\textbf{(ii)} Second, the Gumbel distribution as a universal large-$N$ limit of the
fastest FPT may also yield inaccurate predictions when applied to
biologically relevant amounts of diffusing particles. As $N$
increases, the right tail of the Gumbel distribution becomes
increasingly accurate, but an extremely large $N$ is needed for its
applicability for extended injection. In other words, there exists an
intermediate range of $N$, for which the Gumbel distribution is not
applicable. This limitation originates from an extended injection
because the Gumbel distribution was seen to be quite accurate even for
moderate $N$ in the case of an instantaneous injection (see
Fig.~\ref{fig:H_1d} in App. \ref{sec:figures}). This is consistent with the
above analysis of the mean fFPT, for which the asymptotic formulas
were also much more accurate for the case of instantaneous
injection. Finding an appropriate approximation for the PDF $H_N(t)$
at intermediate values of $N$ presents an interesting open problem.

\section{Discussion}

\subsection*{Approach to the Gumbel distribution}

The distribution of the minimum of a large number $N$ of independent
random variables is known to approach a Gumbel distribution as
$N\to\infty$.  In our setting of the fFPT, it means that
\begin{equation}
H_N(t)\underset{N\to\infty}{\sim}\frac{H^{\rm G}\bigl(\frac{t-b_N}{a_N}
\bigr)}{a_N},\quad H^{\rm G}(x)=\exp(x-e^{x}),
\end{equation}
with appropriate scale and location parameters $a_N$ and $b_N$ (note
that $\exp(-x-e^{-x})$ that is often referred to as the Gumbel
density, corresponds to $H^{\rm G}(-x)$).  For the case of an
instantaneous injection of $N$ particles, Lawley found the asymptotic
behavior of these two parameters under the assumption
(\ref{eq:St0_cond}) on the short-time behavior of the survival
probability $S(t)$ (see \cite{Lawley20} and a summary in
App.~\ref{sec:Lawley}). Despite the universal character of this law,
an approach of $H_N(t)$ to its limiting Gumbel form as $N$ becomes
large can be extremely slow. In this light, it is remarkable how
accurately the Gumbel density $H^{\rm G}$ does approximate $H_N(t)$
even for moderately large $N$ in earlier-studied examples of the
diffusive dynamics under an instantaneous injection (see examples in
\cite{Lawley20}, as well as Fig.~\ref{fig:H_1d} in 
App. \ref{sec:figures}). Conversely, our numerical results from
Sec.~\ref{sec:3} illustrate the opposite---and presumably more
generic---situation of a very slow convergence. Indeed, even though
the injection profile with a power-law behavior at short times does
not change the leading-order terms of $a_N$ and $b_N$, the Gumbel
distribution does not approximate $H_N(t)$ for moderately large $N$,
as shown in Fig.~\ref{fig:H_1d_gamma_nu5}. Moreover, the asymptotic
behavior of $H_N(t)$ turns out to be quite sensitive to the
pre-exponential (subleading) factor and its exponent
$\bar{\alpha}=\alpha+2\nu$. In particular, a faster power-law decay of
the injection profile $\psi(t)$ as $t\to0$ effects larger deviations
from the Gumbel distribution. However, it is instructive to mention
that this empirical observation is not valid in general. To illustrate
this point, let us consider an injection profile of the form
\begin{equation}
\psi(t)=\frac{\sqrt{c}\,e^{-c/t}}{\sqrt{\pi t^3}},
\end{equation}
with the time scale $c>0$.  As this profile vanishes faster than any
power-law as $t\to 0$, one might intuitively expect that the
convergence would be even worse for this profile.  However, it is easy
to check that Eq.~(\ref{eq:Hpsi}) yields
\begin{equation}  
H_\psi(t)=\frac{\sqrt{c}+\sqrt{C}}{\sqrt{\pi t^3}}\,e^{-(\sqrt{c}+\sqrt{C})
^2/t},
\end{equation}
where $C=x_0^2/(4D)$, i.e., we get exactly the same probability
density as for an instantaneous injection, only with the rescaled
parameter $\bar{C}=(\sqrt{c} +\sqrt{C})^2$ instead of $C$. As a
consequence, the approach of $H_N(t)$ to the Gumbel density should be
as rapid as for the instantaneous injection. In summary, the
convergence rate to the Gumbel distribution, as well as the
approximate forms of $H_N(t)$ for moderately large $N$ need to be
better understood.

\subsection*{Random injection}

In the case of random injection, it may be informative to perform the
asymptotic analysis of the survival probability $S_\psi(t)$ in the Laplace
domain. In fact, the short-time behavior of $S_\psi(t)$ can be determined
from the large-$p$ behavior of its Laplace transform,
\begin{equation}
\tilde{S}_\psi(p)=\int\limits_0^\infty dt\,e^{-pt}\,S_\psi(t)=\frac{1-
\tilde{H}_\psi(p)}{p},
\end{equation}
where 
\begin{equation}
\tilde{H}_\psi(p)=\int\limits_0^\infty dt\,e^{-pt}H_\psi(t)=\langle e^{-p
(\tau_k+\delta_k)}\rangle=\tilde{H}(p)\tilde{\psi}(p),
\end{equation}
due to the independence of $\tau_k$ and $\delta_k$. To proceed, it is
convenient to reformulate the condition (\ref{eq:St0_cond}) in the Laplace
domain. Using the identity
\begin{equation}
\int\limits_0^\infty dz\,z^{\nu-1}e^{-c/z-z/b}=2(cb)^{\nu/2}K_\nu(2\sqrt{c/b}),
\end{equation}
where $K_\nu(z)$ is the modified Bessel function of the second kind, we can
evaluate the Laplace transform of Eq.~(\ref{eq:St0_cond}) as
\begin{align}
\nonumber
\frac{1}{p}-\tilde{S}(p)&\sim 2A(C/p)^{(\alpha+1)/2}K_{\alpha+1}\bigl(2
\sqrt{Cp}\bigr)\\
\label{eq:Sp_cond}
&\approx\sqrt{\pi}AC^{\alpha/2+1/4}p^{-\alpha/2-3/4}e^{-2\sqrt{Cp}}\quad(p
\to\infty),
\end{align}
where we used the asymptotic behavior of the modified Bessel function at
large argument. In other words, one can use either the condition
(\ref{eq:St0_cond}) in the time domain or the condition (\ref{eq:Sp_cond})
in the Laplace domain. Note that Eq.~(\ref{eq:Sp_cond}) can also be written
as
\begin{equation}
\label{eq:Hp_cond}
\tilde{H}(p)\sim\sqrt{\pi}AC^{\alpha/2+1/4}p^{-\alpha/2+1/4}e^{-2\sqrt{Cp}}
\quad(p\to\infty).
\end{equation}
In this way, one can further analyze the effect of the injection profile
$\psi(t)$ through the large-$p$ asymptotic behavior of its Laplace
transform $\tilde{\psi}(p)$.

\subsection*{On the short-time limit}

We emphasize the following points. In the present paper, as well as in
the previous works on the instantaneous injection of $N$ particles
into the domain of interest
\cite{Weiss83,Schuss19,Basnayake19,Lawley20,Lawley20b,Lawley20c,Madrid20,Grebenkov20a,Grebenkov22,LawleyBook},
it is stipulated that the particles perform a standard Brownian motion
from time $t=0$. That implies that there is a tacit underlying
assumption that each particle, during an infinitesimally small time
interval $\delta t\to0$ moves an infinitesimally small distance
$\delta a\to0$, with the ratio $D=\delta a^2/(2\delta t)$ kept
fixed. Such an assumption is conventional and valid for many
practically important situations. Concurrently, as it was demonstrated
here as well as in the previous works, in the limit $N\to\infty$ the
corresponding asymptotic forms of the survival probability of the
target and the ensuing fFPT-PDF become fully dominated by the behavior
of the process in a very short time-span in the vicinity of $t=0$.  In
the limit $N\to \infty$, this span becomes so small that a particle
can make, at most, a few jumps away from its initial location, and its
short-time dynamics in realistic systems may be very different from
that of a conventional (continuous) Brownian motion. Indeed, for the
latter a particle may travel on any distance within an arbitrarily
short time interval with an exponentially small but non-zero
probability, while in realistic settings this probability should be
identically equal to zero up to a certain time instant.\footnote{Such
a finite-horizon behavior is, e.g., essential in the description of
heat conduction dynamics, for which the diffusion equation is
typically replaced by the hyperbolic telegrapher's or Cattaneo
equation, endowing the process with a finite propagation speed
\cite{jou,cattaneo}.}  This signifies, in particular, that the
logarithmic reduction of the mean first-passage time in
Eq.~\eqref{weiss}, as well as our results presented here are only
valid for sufficiently large but still
\textit{bounded\/} values of $N$, such that the time-interval which
dominates the behavior of the survival probability should remain
sufficiently large permitting to use the Brownian motion picture.

The true asymptotic behavior of the characteristic properties in the
mathematical limit $N\to\infty$ may appear somewhat different, but to
determine it one would need to start with a discrete-space approach which
captures the small nuances of the short-time behavior, especially in the
case of processes with a broad distribution of waiting times, which
ultimately result in a large-scale subdiffusive motion. Such a robust
approach is lacking at present but clearly represents a challenging and
pressing issue.

\section{Conclusions}
\label{sec:5}
 
In summary, we investigated the effect of an extended, time-dependent
injection (or entrance) of particles on the statistics of the fastest
arrival $\T_N$ to the target. While a formal implementation of
prescribed time delays $\delta_k$ is straightforward for independently
diffusing particles, the asymptotic analysis of the PDF $H_N(t)$ of
the fastest FPT $\T_N$ and its mean $\langle\T_N\rangle$ at large $N$
required a substantial reformulation of the problem. The first
deterministic formulation was based on a continuous approximation of
the injection profile $\psi(t)$ and allowed us to introduce effective
first-passage times $\bar{\tau}_k$ that account for a given profile
through the modified survival probability $\bar{S}_\psi(t)$ given by
Eq.~(\ref{Spsibar}). The second probabilistic formulation treated the
delay times $\delta_k$ as independent random variables generated from
the profile $\psi(t)$ so that the delayed FPTs $\tau_k+\delta_k$ were
determined by the survival probability $S_\psi(t)$ given by
Eq.~(\ref{Spsi}). We showed that both formulations yielded remarkably
similar results, even for a single particle (see
Fig.~\ref{fig:Hprime}). 

Next, we analyzed the impact of the injection
profile $\psi(t)$ onto the survival probability $S_\psi(t)$. Under
rather general assumptions on the short-time behaviors of $S(t)$ and
$\psi(t)$, we derived the short-time behavior of $S_\psi(t)$ and thus
the large-$N$ asymptotic behavior (\ref{eq:meanFPT_asympt0}) of the
mean fFPT. This is an extension of former results known for an instantaneous
injection. We also showed that the extended injection considerably
slows down the approach of the mean fFPT to its asymptotic value in
the limit $N\to\infty$. Similarly, we showed that an approach of
$H_N(t)$ to its limiting Gumbel distribution upon an increase of $N$
is also very slow, such that the Gumbel distribution may not be
applicable for an intermediate range of $N$ from tens to thousands,
which is the most relevant scenario in molecular biology. This result
urges for further investigations of this fundamental problem and
obtaining correction terms to the Gumbel distribution in this setting.

\begin{acknowledgments}
DSG acknowledges the Alexander von Humboldt Foundation for support
within a Bessel Prize award. RM acknowledges funding from the German
Science Foundation (DFG, grant ME 1535/22-1 and ME 1535/16-1).
\end{acknowledgments}

\appendix

\section{Survival probability $\bar{S}_\psi(t)$}
\label{sec:auxil}

In this Appendix, we discuss some properties of the survival
probability $\bar{ S}_\psi(t)$ defined by Eq.~(\ref{Spsibar}). We here
assume that the survival probability $S(t)$ vanishes as $t\to\infty$
(as it is always the case for diffusion in a bounded domain) and aim
at showing that $\bar{S}_\psi(t)$ monotonously decreases to $0$ as $t$
increases. For convenience, we introduce the notation
\begin{equation}
\label{eq:fpsi}
f_\psi(t)=-\int\limits_0^t dt'\,\psi(t')\,\ln S(t-t').
\end{equation}

First, Jensen's inequality for the concave function $\ln(z)$ implies
\begin{align}
\nonumber
&\frac{1}{1-\Psi(t)}\int\limits_0^tdt'\,\psi(t')\,\ln S(t-t')\\
\label{eq:auxil1}
&\leq\ln\left(\frac{1}{1-\Psi(t)}\int\limits_0^tdt'\,\psi(t')\,S(t-t')\right),
\end{align}
where
\begin{equation}
\Psi(t)=\int\limits_t^\infty dt'\,\psi(t').
\end{equation}
Second, the integral on the right-hand side can be split into the two
contributions
\begin{align}
\nonumber
I(t)&=\int\limits_0^tdt'\,\psi(t')\,S(t-t')\\
&=\int\limits_0^{t/2}dt'\,\psi(t')\,S(t-t')+\int\limits_0^{t/2}dt'\,\psi(t-t')
\,S(t'),
\end{align}
where the integration variable $t'$ was changed to $t-t'$ in the second
integral. As the survival probability $S(t)$ is a monotonously decreasing
function, one can estimate the first integral as
\begin{equation}
\int\limits_0^{t/2}dt'\,\psi(t')\,S(t-t')\leq S(t/2)\int\limits_0^{t/2}dt'\,
\psi(t')\leq S(t/2),
\end{equation}
where the last integral was extended to $+\infty$ and then replaced by unity
due to the normalization of $\psi(t)$. Since $S(t)$ vanishes in the limit $t
\to\infty$, this contribution vanishes. Similarly, we estimate the second
integral as
\begin{equation}
\int\limits_0^{t/2}dt'\,\psi(t-t')\,S(t')\leq\int\limits_0^{t/2}dt'\,\psi(t-t')
=\Psi(t/2)-\Psi(t).
\end{equation}
In the long-time limit, $\Psi(t)$ vanishes and thus does the second integral.
We conclude that $I(t)\to 0$ as $t\to\infty$ so that the right-hand side of
Eq.~(\ref{eq:auxil1}) goes to $-\infty$ and thus $f_\psi(t)$ from
Eq.~(\ref{eq:fpsi}) diverges to $+\infty$, whereas $\bar{S}_\psi(t)=e^{-f_
\psi(t)}$ vanishes.

Note that the inequality (\ref{eq:auxil1}) also implies that
\begin{align}
\nonumber
\bar{S}_\psi(t)&\leq\exp\left((1-\Psi(t))\ln\left(\frac{\int\nolimits_0^tdt'
\,\psi(t')\,S(t-t')}{1-\Psi(t)}\right)\right)\\
\label{eq:ineq1}
&=\left(1-\frac{1-S_\psi(t)}{1-\Psi(t)}\right)^{1-\Psi(t)},
\end{align}
where we used the definition (\ref{Spsi}) of $S_\psi(t)$. Note that
Eq.~(\ref{Spsi}) implies that $S_\psi(t)\geq\Psi(t)$, i.e., the right-hand
side is nonnegative. Re-arranging the above inequality, one gets the
equivalent form
\begin{equation}
S_\psi(t)\geq\Psi(t)+(1-\Psi(t))[\bar{S}_\psi(t)]^{1/(1-\Psi(t))}.
\end{equation}
Since $(1-x/a)^a\leq1-x$ for any $x\in(0,a)$ and any $0<a\leq1$, one can also
deduce from the inequality (\ref{eq:ineq1}) the simpler bound
\begin{equation}
\bar{S}_\psi(t)\leq S_\psi(t).
\end{equation}
The probabilistic interpretation of this inequality is that the
first-passage time $\ttau_k$ is typically smaller than
$\tau_k+\delta_k$, i.e., the deterministic injection is faster than
the random one (for the same profile $\psi(t)$).

\section{Short-time behavior of the survival probability}
\label{sec:short}

In this Appendix, we derive the short-time asymptotic behavior of the survival
probabilities $S_\psi(t)$ and $\bar{S}_\psi(t)$. 

Rewriting Eq.~(\ref{Spsi}) as
\begin{equation}  
1-S_\psi(t)=\int\limits_0^tdt'\,\psi(t-t')(1-S(t'))
\end{equation}
and substituting the asymptotic relations (\ref{eq:St0_cond}) and
(\ref{eq:psi_short}), we get
\begin{align}
\nonumber
&1-S_\psi(t)\approx\int\limits_0^tdt'\,a(t-t')^{\nu-1}e^{-c^\mu/(t-t')^\mu}\,
A(t')^\alpha e^{-C/t'}\\
&=aAt^{\nu+\alpha}\int\limits_0^1dz\,z^\alpha\,(1-z)^{\nu-1}\,e^{-(c/t)^\mu
(1-z)^{-\mu}-(C/t)/z},
\end{align}
where we changed the integration variable to $t'=tz$. To estimate the asymptotic
behavior of this integral in the limit $t\to0$, we write the exponential
function as $e^{-(c/t)^\mu f(z)}$, with
\begin{equation}
f(z)=\frac{1}{(1-z)^\mu}+\frac{\mu B}{z},\quad B=\frac{Ct^{\mu-1}}{\mu c^\mu} 
\end{equation}
and apply the Laplace method. For this purpose, one determines $z_0$ as the
minimum of the function $f(z)$, given by the equation
\begin{equation}
\label{eq:fprime_z0}
f'(z_0)=\frac{\mu}{(1-z_0)^{\mu+1}}-\frac{\mu B}{z_0^2}=0.
\end{equation}
As a consequence, the integral reads
\begin{eqnarray}
\nonumber
1-S_\psi(t)&\approx&aAt^{\nu+\alpha}e^{-(c/t)^\mu f(z_0)}z_0^\alpha(1-z_0)^{
\nu-1}\\
&&\times\sqrt{\frac{2\pi t^\mu}{c^\mu f''(z_0)}}.
\end{eqnarray}

We consider the following four cases:

(i) For $\mu>1$, the factor $(c/t)^{\mu}$ is dominant. At very short times $t$,
one has $B\ll1$, so that $z_0$ is small and the above equation can be solved as
$z_0\approx\sqrt{B}\ll1$, and thus $f(z_0)\approx1+2\mu\sqrt{B}$ and $f''(z_0)
\approx\mu(\mu+1)+2\mu/\sqrt{B}$. To leading order, we get
\begin{eqnarray}
\nonumber
1-S_\psi(t)&\approx&aAt^{\nu+ \mu(2\alpha+3)/4-1/4}\,e^{-(c/t)^\mu}\\
&&\times(C/(\mu c^\mu))^{\alpha/2+1/4}\sqrt{\frac{\pi}{c^\mu\mu}}.
\end{eqnarray}
We therefore find the asymptotic behavior (\ref{eq:St_cond}) with 
\begin{subequations}
\label{eq:Abar_case1}
\begin{align}
\bar{C}&=c,\quad\bar{\mu}=\mu,\\
\bar{\alpha}&=\nu+\mu(2\alpha+3)/4-1/4,\\
\bar{A}&=aA(C/(\mu c^\mu))^{\alpha/2+3/4}\sqrt{\pi/C}.
\end{align} 
\end{subequations}

(ii) For $\mu=1$, one can solve Eq.~(\ref{eq:fprime_z0}) to get $z_0=\sqrt{B}/
(1+\sqrt{B})$, where $B=C/c$ is independent of $t$. One then finds $f(z_0)=(1
+\sqrt{B})^2$ and $f''(z_0)=2(1+1/\sqrt{B})$, from which
\begin{eqnarray}
\nonumber
1-S_\psi(t)&\approx&aAt^{\nu+\alpha+1/2}e^{-(\sqrt{c}+\sqrt{C})^2/t}\\
&&\times\frac{B^{\alpha/2+1/4}}{(1+\sqrt{B})^{\alpha+\nu+1/2}}\sqrt{\pi/c}.
\end{eqnarray}
In other words, we retrieve the asymptotic behavior (\ref{eq:St_cond}) with
the parameters
\begin{subequations}
\label{eq:Abar_case2}
\begin{align}
\bar{C}&=(\sqrt{c}+\sqrt{C})^2,\quad\bar{\mu}=1,\\
\bar{\alpha}&=\alpha+\nu+1/2,\\
\bar{A}&=aA\frac{(C/c)^{\alpha/2+1/4}}{(1+\sqrt{C/c})^{\alpha+\nu+1/2}}\sqrt{
\pi/c}.
\end{align}
\end{subequations}

(iii) For $0<\mu<1$, one has $B\gg1$ in the short-time limit, so that the
solution of Eq.~(\ref{eq:fprime_z0}) is approximately $z_0\approx1-B^{-1/(1+
\mu)}$ and thus $f(z_0)\approx t^{\mu-1}C/c^\mu$ and $f''(z_0)=\mu(\mu+1)B^{
(\mu+2)/(\mu+1)}+2\mu B$, such that
\begin{eqnarray}
\nonumber
1-S_\psi(t)&\approx&aAt^{\alpha+(2\nu+\mu)/(\mu+1)}e^{-C/t}\\
&&\hspace*{-1.8cm}\times(C/(\mu c^\mu))^{-(\nu+\mu/2)/(\mu+1)}\sqrt{\frac{
2\pi}{c^\mu\mu(\mu+1)}}.
\end{eqnarray}
Once again, we find Eq.~(\ref{eq:St_cond}), with
\begin{subequations}
\label{eq:coeff_mu0_1}
\begin{align}
\bar{C}&=C,\quad\bar{\mu}=1,\\
\bar{\alpha}&=\alpha+(2\nu+\mu)/(\mu+1),\\
\label{eq:A_case3}
\bar{A}&=aA(C/(\mu c^\mu))^{-(\nu+\mu/2)/(\mu+1)}\sqrt{\frac{2\pi}{c^\mu\mu(
\mu+1)}}.
\end{align}
\end{subequations}

(iv) The case $\mu=0$ requires a separate analysis because Eq.~(\ref{eq:A_case3})
either vanishes or diverges as $\mu\to0$. Here one has to distinguish the cases
$\nu>1$ and $\nu\leq1$. We discuss only the former case when the injection
profile vanishes at $t=0$, for which
\begin{equation}
1-S_\psi(t)\approx aAt^{\alpha+\nu}\int\limits_0^1dz\,\exp(-f(z)),
\end{equation}
with $f(z)=C/(tz)-\alpha\ln z-(\nu-1)\ln(1-z)$. The minimum $z_0$ of this
function via
\begin{equation}
f'(z_0)=-\frac{C}{tz_0^2}-\frac{\alpha}{z_0}+\frac{\nu-1}{1-z_0}=0.
\end{equation}
As $C/t$ is large, the first term should be compensated by the third term, such
that $z_0$ should be close to $1$. Substituting $z_0=1-\ve$ into this equation
and expanding up to linear order in $\ve$, we get $\ve\approx(\nu-1)/(C/t+
\alpha+2(\nu-1))$. To leading order, we also get
\begin{equation}
f''(z_0)= \frac{2C}{tz_0^3}+\frac{\alpha}{z_0^2}+\frac{\nu-1}{(1-z_0)^2}
\approx\frac{(t/C)^2}{\nu-1}.
\end{equation}
The Laplace method yields again Eq.~(\ref{eq:St_cond}) to leading order, with
\begin{subequations}
\label{eq:coeff_mu0}
\begin{align}
\bar{C}&=C,\quad\bar{\mu}=1,\\  
\bar{\alpha}&=\alpha+2\nu,\\
\bar{A}&=aAC^{-\nu}\,\bigl[e^{1-\nu}(\nu-1)^\nu\sqrt{2\pi/(\nu-1)}\bigr]\\
&\approx aAC^{-\nu}\Gamma(\nu). 
\end{align}
\end{subequations}
For the last relation, we identified the factor in the square brackets as the
Stirling expansion of the Gamma function $\Gamma(\nu)$ for $\nu\gg1$. Note
that the above estimation gets more accurate for large $\nu$.  

While we focused on the analysis of the survival probability $S_\psi(t)$, the
above results are also applicable to $\bar{S}_\psi(t)$. In fact, one has
\begin{align}
\nonumber
&1-\bar{S}_\psi(t)=1-\exp\left(\int\limits_0^tdt'\psi(t-t')\ln(1-(1-S(t')))
\right)\\
&\approx\int\limits_0^tdt'\,\psi(t-t')(1-S(t'))=1-S_\psi(t)\quad(t\to 0),
\end{align}
where the logarithm was expanded into a Taylor series in powers of $(1-S(t'))$,
which is small at short times. Note that the correction term includes $(1-S_
\psi(t))^2$ and an integral involving $(1-S(t'))^2$, which are both
exponentially small as compared to $1-S_\psi(t)$.

\section{Primer on the asymptotic results for instantaneous injection}
\label{sec:Lawley}

In the large $N$ limit, the fastest FPT $\T_N^0=\min\{\tau_1,\ldots,\tau_N\}$
for the case of instantaneous injection converges to the Gumbel distribution
with the scale and location parameters $a_N$ and $b_N$, i.e., $(\T_N^0-b_N)/
a_N$ converges in distribution to a random variable $X$ obeying the standard
Gumbel distribution $\P\{X>x\}=\exp(-e^x)$ on $\R$. Under the assumption
(\ref{eq:St0_cond}) on the short-time behavior of the survival probability
$S(t)$, Lawley derived two equivalent asymptotic forms for the parameters
$a_N$ and $b_N$ \cite{Lawley20}. The first form is more accurate but less
explicit,
\begin{align}  
a_N&=\frac{b_N}{\ln(AN)},\qquad b_N=\frac{C}{\ln(AN)}\qquad(\alpha=0),\\
\label{eq:aNbN_Lawley}
a_N&=\frac{b_N}{\alpha(1+W)},\qquad b_N=\frac{C}{\alpha W}\qquad(\alpha\ne0), 
\end{align}
and
\begin{equation}
W=\left\{\begin{array}{ll}W_0\bigl(C(AN)^{1/\alpha}/\alpha\bigr)&(\alpha
>0),\\[0.2cm]
W_{-1}\bigl(C(AN)^{1/\alpha}/\alpha\bigr)&(\alpha<0),\end{array}\right.
\end{equation}
where $W_0(z)$ denotes the principal branch of the Lambert W-function and
$W_{-1}(z)$ denotes the lower branch \cite{Corless96}. The second form
(denoted by a prime) can be deduced from the asymptotic behavior of
the Lambert function \cite{Lawley20},
\begin{equation}
a'_N=\frac{C}{(\ln N)^2},\quad b'_N=\frac{C}{\ln N}+\frac{C[\alpha\ln\ln N-
\ln(AC^\alpha)]}{(\ln N)^2}.
\end{equation}
In particular, the mean and variance of $\T^0_N$ (whenever they exist) are
\begin{align}
\label{eq:TN0_mean0}
\langle\T_N^0\rangle&\approx b_N-a_N\gamma+o(a_N),\\
\label{eq:TN0_mean}
&\approx\frac{C}{\ln N}+C\frac{\alpha\ln\ln N-\ln(A C^\alpha)-\gamma}{(\ln N)^2}
\end{align}
and
\begin{align}
\label{eq:TN0_var}
\mathrm{Var}\{\T_N^0\}&\approx\frac{\pi^2}{6}a_N^2+o(a_N^2)\\
&\approx\frac{\pi^2}{6}\,\frac{C^2}{(\ln N)^4},
\end{align}
where $\gamma\approx0.5772$ is the Euler constant. Note also that the most
probable time of the Gumbel distribution is precisely given by $b_N$, i.e.,
\begin{equation}
\T_{N,{\rm mp}}^0\approx b_N\approx\frac{C}{\ln N}+O(1/(\ln N)^2).
\end{equation}

\begin{figure}
\begin{center}
\includegraphics[width=80mm]{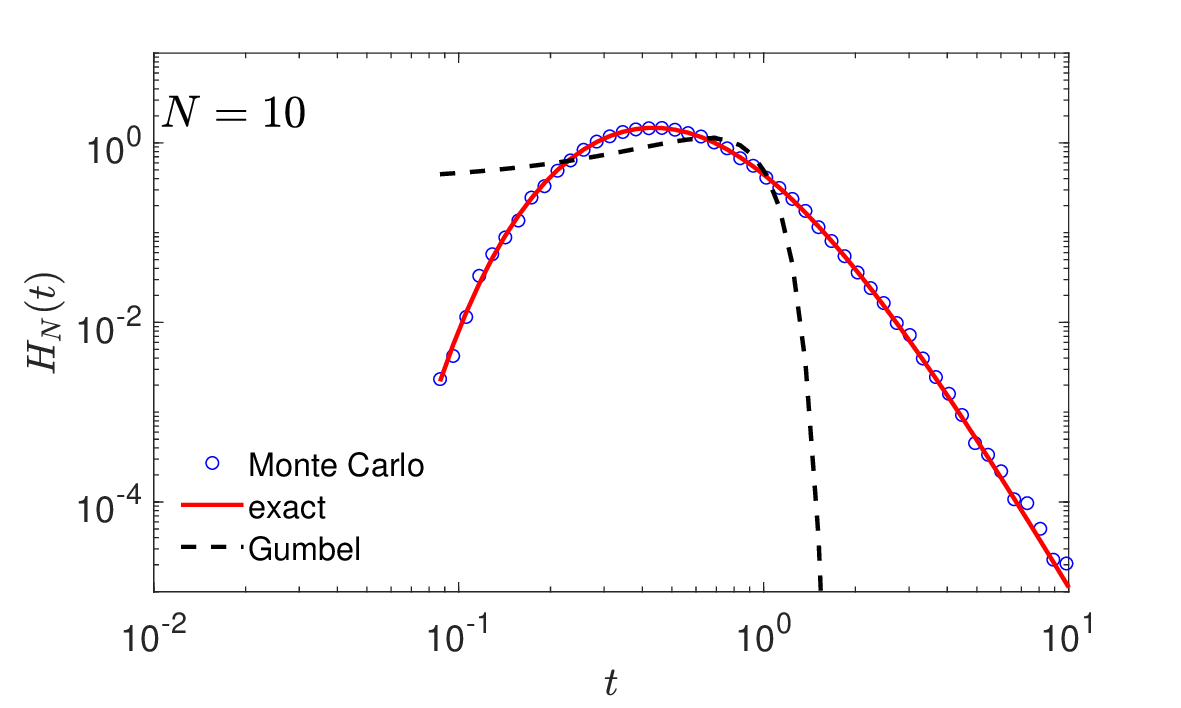} 
\includegraphics[width=80mm]{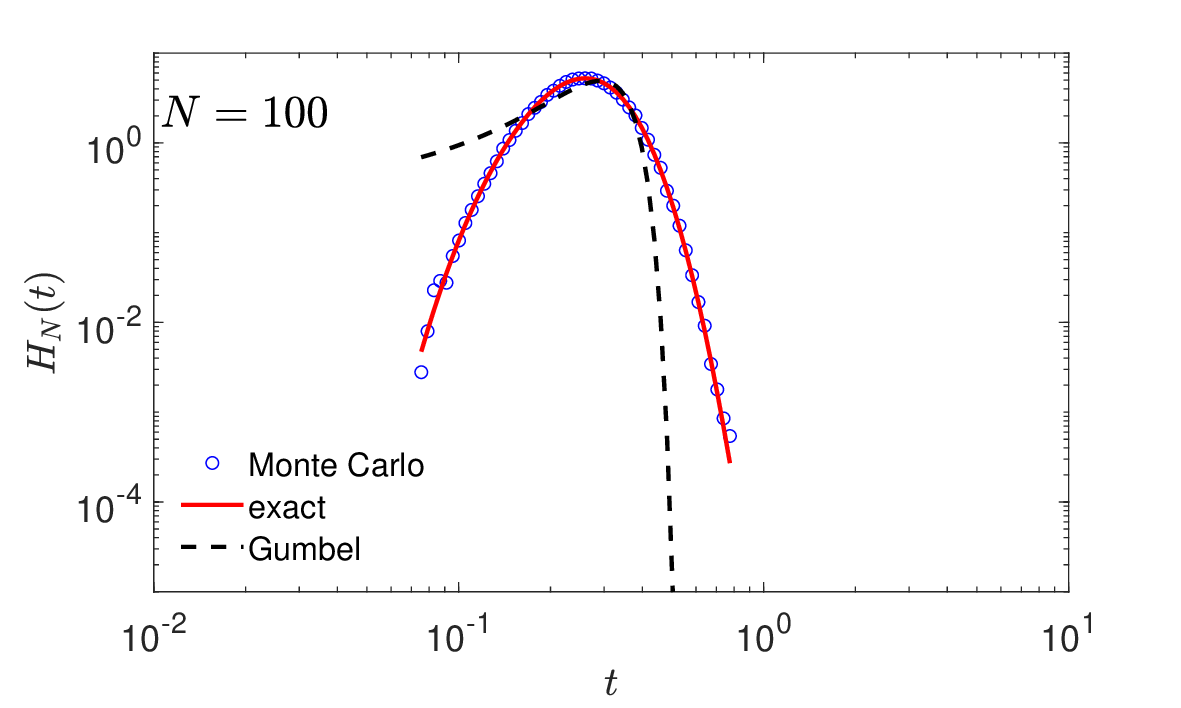} 
\includegraphics[width=80mm]{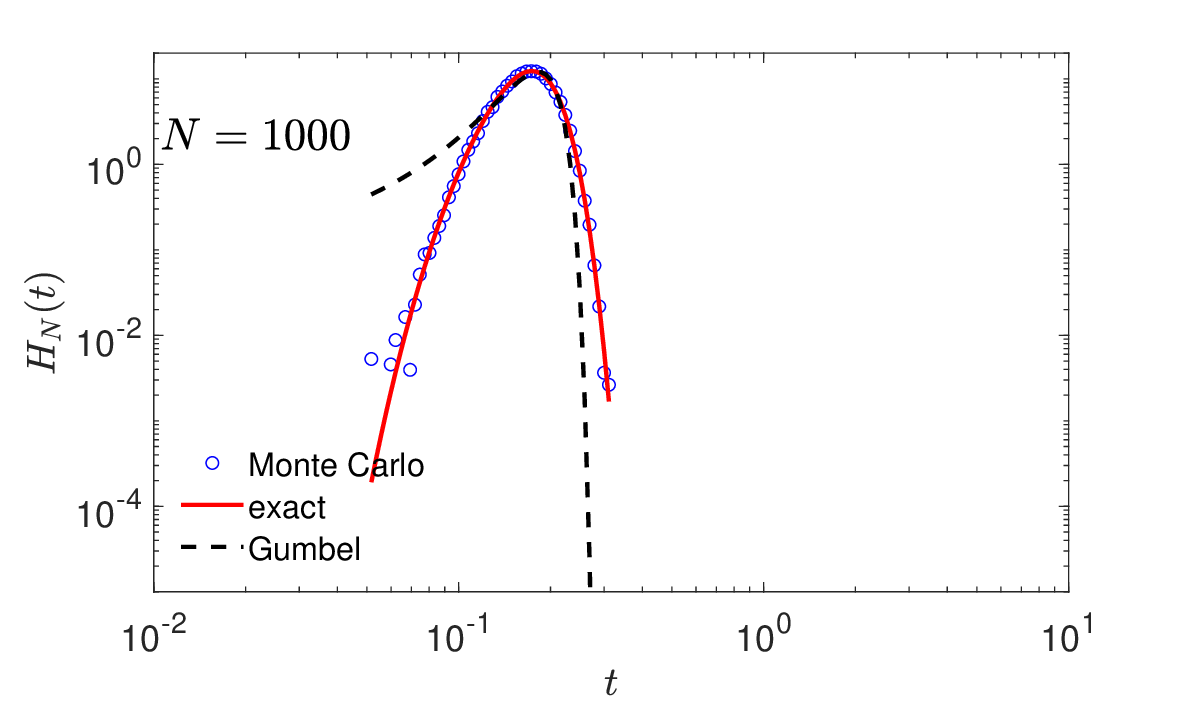} 
\includegraphics[width=80mm]{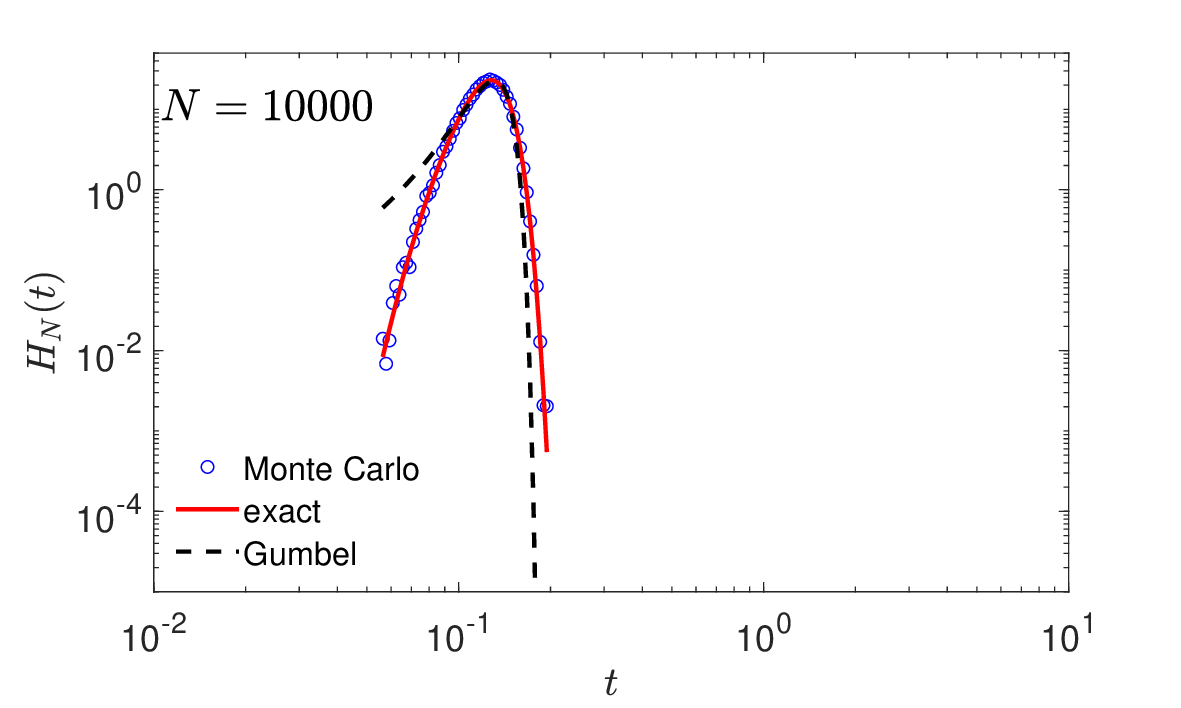} 
\end{center}
\caption{PDF $H_N^0(t)$ of the fastest FPT $\T_N^0$ for diffusion on the
half-line, with $x_0=2$, $D=1$, and an instantaneous injection of $N$
particles, for four values of $N$ (from top to bottom): $N=10$, $N=100$,
$N=1000$, and $N=10000$. The circles represent rescaled histograms obtained
by generating $\T_N^0$ (with $10^5$ realizations); the solid line gives
the exact solution $H_N^0(t)=NH(t)[S(t)]^{N-1}$; the dashed line shows the
asymptotic Gumbel distribution, with the parameters $a_N$ and $b_N$ given by
Eq.~(\ref{eq:aNbN_Lawley}) derived by Lawley \cite{Lawley20}.}
\label{fig:H_1d}
\end{figure}

Figure \ref{fig:H_1d} shows the PDF $H_N(t)$ for an instantaneous injection of
$N$ particles diffusing in the half-line. One sees how the distribution slowly
approaches the asymptotic Gumbel distribution. We recall that both the mean
fFPT $\langle\T_N^0\rangle\simeq C/\ln(N)$ and the coefficient of variation,
$\mathrm{Std}\{ \T_N^0\}/\langle \T_N^0\rangle\simeq(\pi/\sqrt{6})/\ln(N)$,
slowly decrease with $N$. As a consequence, the PDF shifts to the left (to
shorter times) and gets narrower as $N$ increases. Note also that the typical
value decreases, as well.

\section{Uniform entrance times}
\label{sec:uniform}

We here briefly discuss another common setting when the entrance times
$\delta_k$ are uniformly distributed over a time span from $0$ to $T$. This
injection profile,
\begin{equation}
\label{eq:psi_uniform}
\psi(t)=\frac{\Theta(T-t)}{T}, 
\end{equation}
in Laplace domain yields $\tilde{\psi}(p)=(1-e^{-pT})/(pT)$, that allows one
to evaluate the survival probability $S_\psi(t)$ explicitly in the case of
diffusion on the half-axis, namely,
\begin{eqnarray}
\nonumber
S_\psi(t)&=&1-\frac{t}{T}F\left(x_0/\sqrt{4Dt}\right)\\
\label{eq:Spsi_uniform}
&&-\frac{t-T}{T}\Theta(t-T)F\left(x_0/\sqrt{4D(t-T)}\right),
\end{eqnarray}
where
\begin{equation}
F(z)=(1+2z^2)\erfc(z)-\frac{2}{\sqrt{\pi}}ze^{-z^2} .
\end{equation}
At short times, the last term in Eq.~(\ref{eq:Spsi_uniform}) is zero
due to the Heaviside step function. In turn, the second term can be
simplified by using $F(z)\approx e^{-z^2}/(\sqrt{\pi}z^3)$ as $z=
x_0/\sqrt{4Dt} \to\infty$, so that we retrieve the asymptotic behavior
(\ref{eq:St_cond}), with
\begin{equation}
\label{eq:param_uniform}
\bar{\mu}=\mu=1,\quad\bar{C}=C=\frac{x_0^2}{4D},\quad\bar{\alpha}=\frac{5}{2},
\quad\bar{A}=\frac{1}{\sqrt{\pi}\,T\,C^{3/2}}.
\end{equation} 
As the uniform profile (\ref{eq:psi_uniform}) exhibits the short-time asymptotic
behavior (\ref{eq:psi_short}) with $\nu=1$ and $\mu=0$, the parameters $\bar{
\alpha}$ and $\bar{A}$ agree with our general expressions (\ref{eq:coeff_mu0}).
Note also that
\begin{eqnarray}
\nonumber
H_\psi(t)&=&\frac{1}{T}\erfc\left(x_0/\sqrt{4Dt}\right)\\
\label{eq:Hpsi_uniform}
&&-\frac{\Theta(t-T)}{T}\erfc\left(x_0/\sqrt{4D(t-T)}\right).
\end{eqnarray}

\begin{figure}
\begin{center}
\includegraphics[width=85mm]{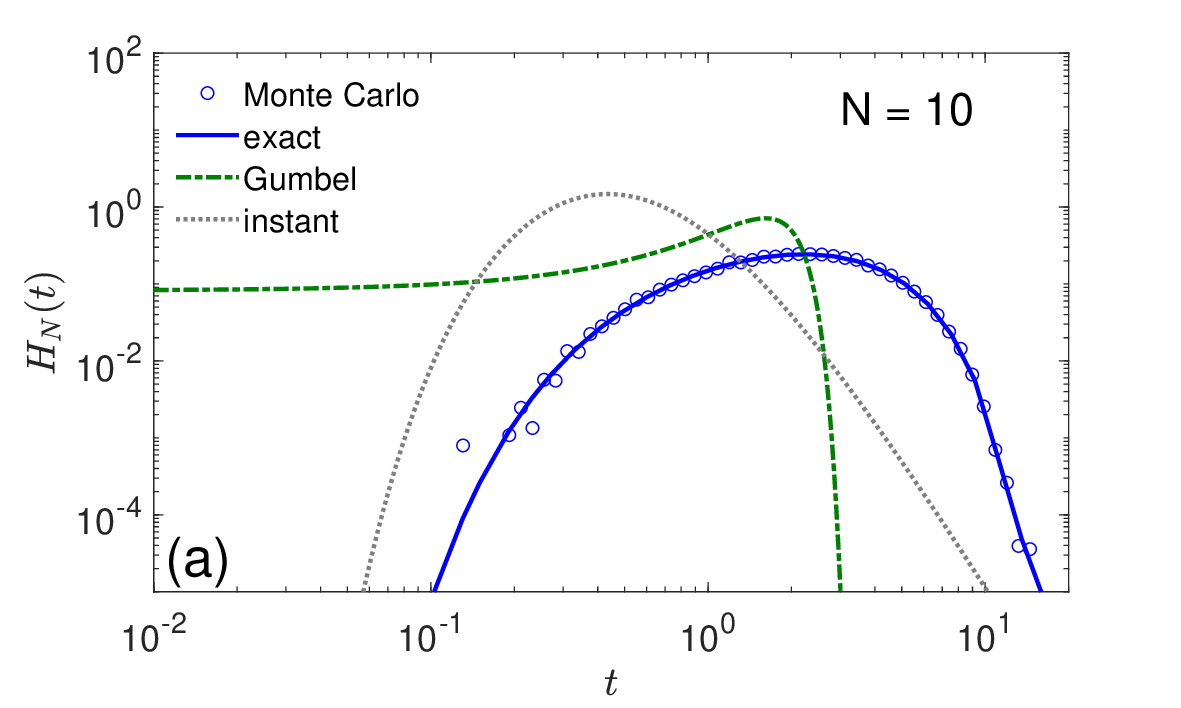} 
\includegraphics[width=85mm]{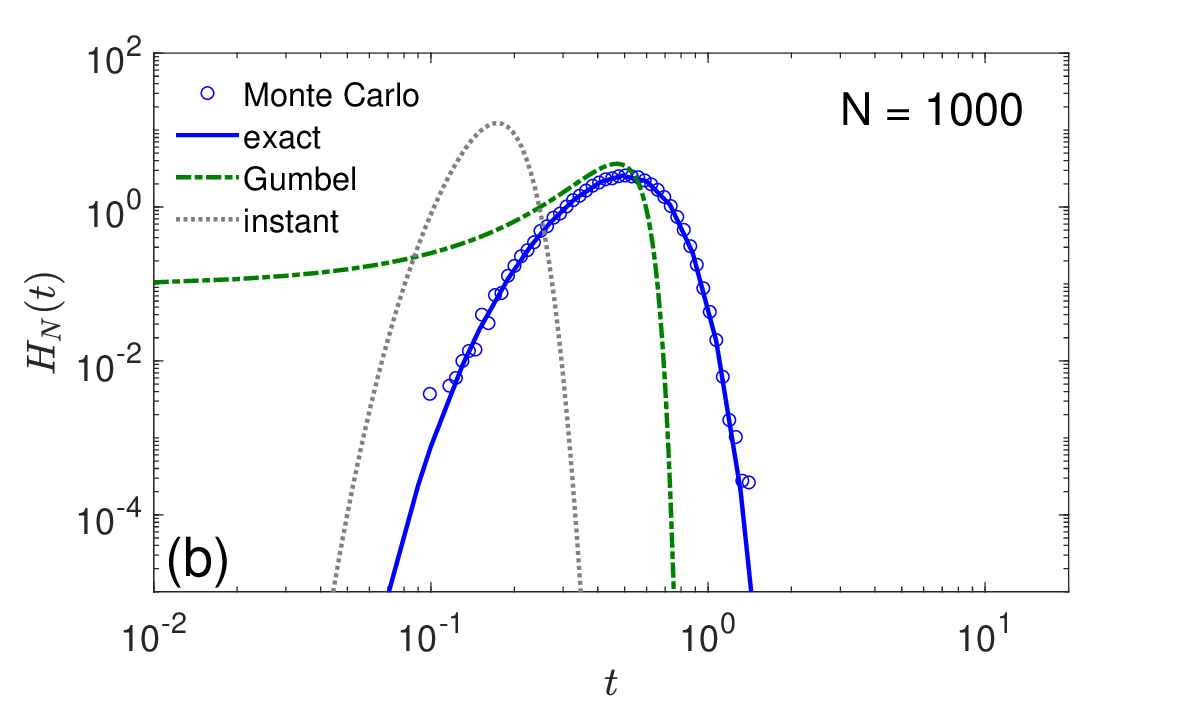} 
\includegraphics[width=85mm]{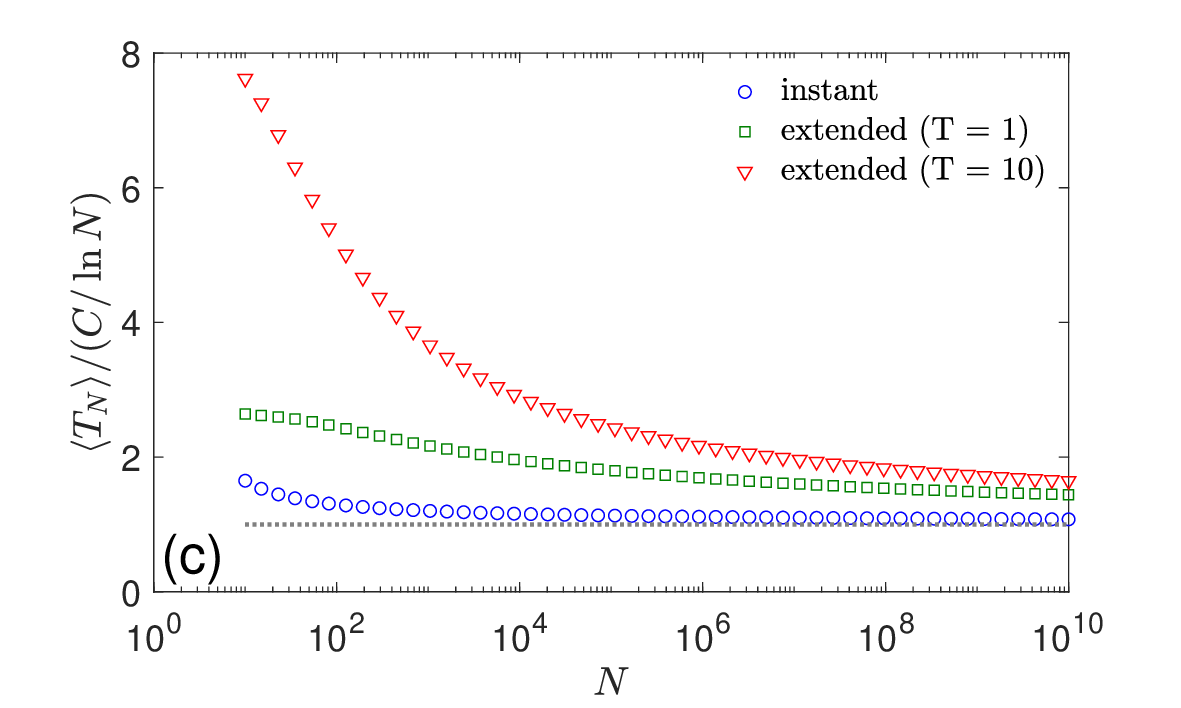} 
\end{center}
\caption{{\bf (a,b)} PDF $H_N(t)$ of the fastest FPT $\T_N$ for diffusion
on the half-line with $x_0=2$ and $D=1$, and an extended injection modeled by
the uniform distribution (\ref{eq:psi_uniform}) with $T=10$, for {\bf (a)}
$N=10$ and {\bf (b)} $N=1000$. The circles represent rescaled histograms
obtained by generating $\T_N$ (with $10^5$ realizations). The solid line
represents the exact solution $H_N(t)=NH_{\psi}(t)[S_{\psi}(t)]^{N-1}$, where
$S_{\psi}(t)$ and $H_{\psi}(t)$ are given by Eqs. (\ref{eq:Spsi_uniform}) and
(\ref{eq:Hpsi_uniform}). The dash-dotted line represents the asymptotic Gumbel
distribution, with the scale and location parameters $a_N$ and $b_N$ given by
Eq.~(\ref{eq:aNbN_Lawley}) with the parameters in Eq.~(\ref{eq:param_uniform}).
The dotted line indicates the exact density for an instantaneous injection.
{\bf (c)} Mean fFPT $\langle\T_N\rangle$, rescaled by $C/\ln N$, as a function
of $N$ for diffusion on the half-line, with $x_0=2$ and $D=1$, for an
instantaneous injection (circles) and an extended injection of $N$ particles
modeled by the uniform distribution (\ref{eq:psi_uniform}) with $T=1$ (squares)
and $T=10$ (triangles). The mean fFPTs were obtained from numerically computing
the integral in Eq.~(\ref{mean2}). The dotted line indicates the constant $1$
to highlight the approach to the leading-order term $C/\ln N$.}
\label{fig:uniform}
\end{figure}

Figure \ref{fig:uniform} illustrates the main features of the fFPT $\T_N$ for
the uniform injection profile: panels (a) and (b) present the PDF $H_N(t)$ for
$T=10$, whereas panel (c) shows the mean fFPT as a function of $N$ for $T=1$
and $T=10$. It is instructive to compare these results to those presented in
the main text for the injection modeled by a gamma distribution. For a proper
comparison, we impose that the mean delay time, $\langle\delta_k\rangle$, is
the same in both cases---under this condition, the uniform distribution
with $T=10$ can be compared to the gamma distribution with $b=1$ and $\nu=5$.
Qualitatively, the PDFs for both cases look similar (compare panels (a,b) with
panels (b,d) of Fig.~\ref{fig:H_1d_gamma_nu5}), but the density $H_N(t)$ is
shifted to longer times for the gamma profile. This is consistent with the
observation that the mean fFPT (shown in Fig.~\ref{fig:Tmean_1d_gamma}) is
longer for the gamma profile, and it approaches slower to the leading-order
term $C/\ln N$. We emphasize that the leading-order term $C/\ln N$ is the
same for both injection profiles such that the actual shape of the profile
influences only the sub-leading term in Eq.~(\ref{eq:meanFPT_asympt0}), as
well as the accuracy of this asymptotic relation.

Panel (c) of Fig.~\ref{fig:uniform} also highlights that the longer mean
delay time $\langle\delta_k\rangle$ (here, $T/2$), expectedly, increases the
mean fFPT $\langle\T_N\rangle$, but its effect strongly depends on the number
$N$ of particles. Moreover, it is in general difficult to distinguish the
relative roles of the mean delay time and the shape of its distribution, as
both affect the subleading term in Eq.~(\ref{eq:meanFPT_asympt0}) via the
modified parameters $\bar{\alpha}$ and $\bar{A}$. Our theoretical
description allows one to analyze these effects for various injection
profiles.

\section{Supplementary illustration}
\label{sec:figures}

In this Appendix, we provide a supplementary illustration to the numerical
results presented in the main text.

\begin{figure}
\begin{center}
\includegraphics[width=85mm]{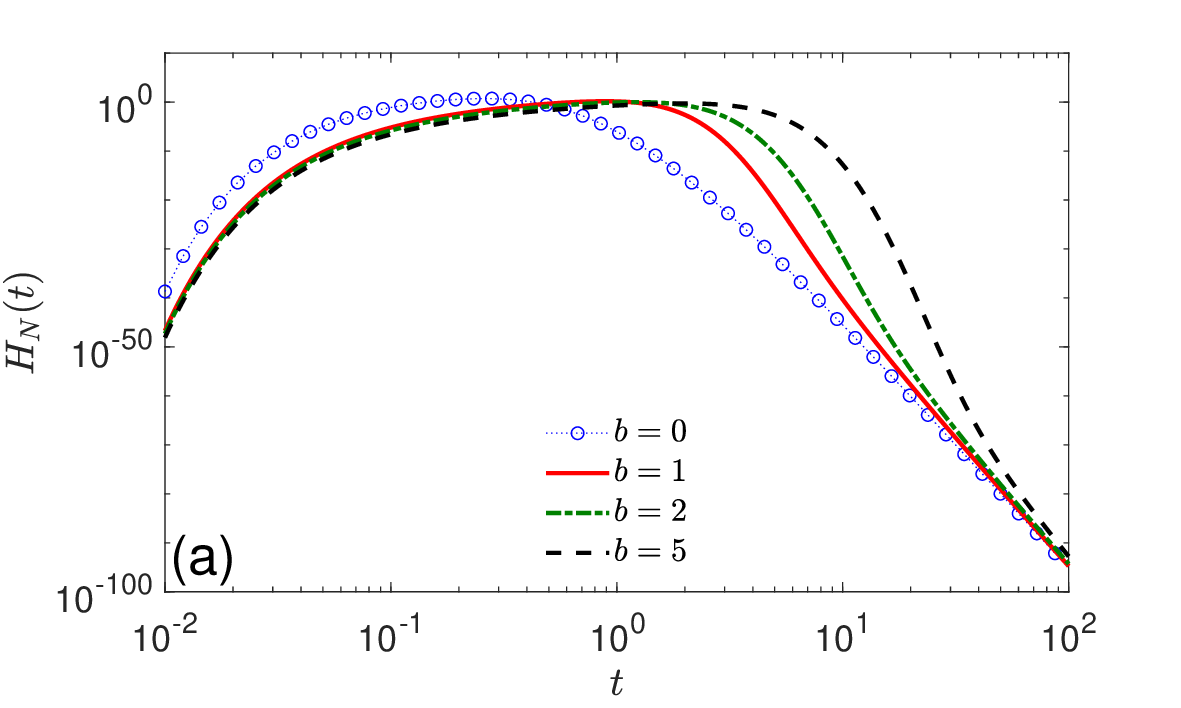} 
\includegraphics[width=85mm]{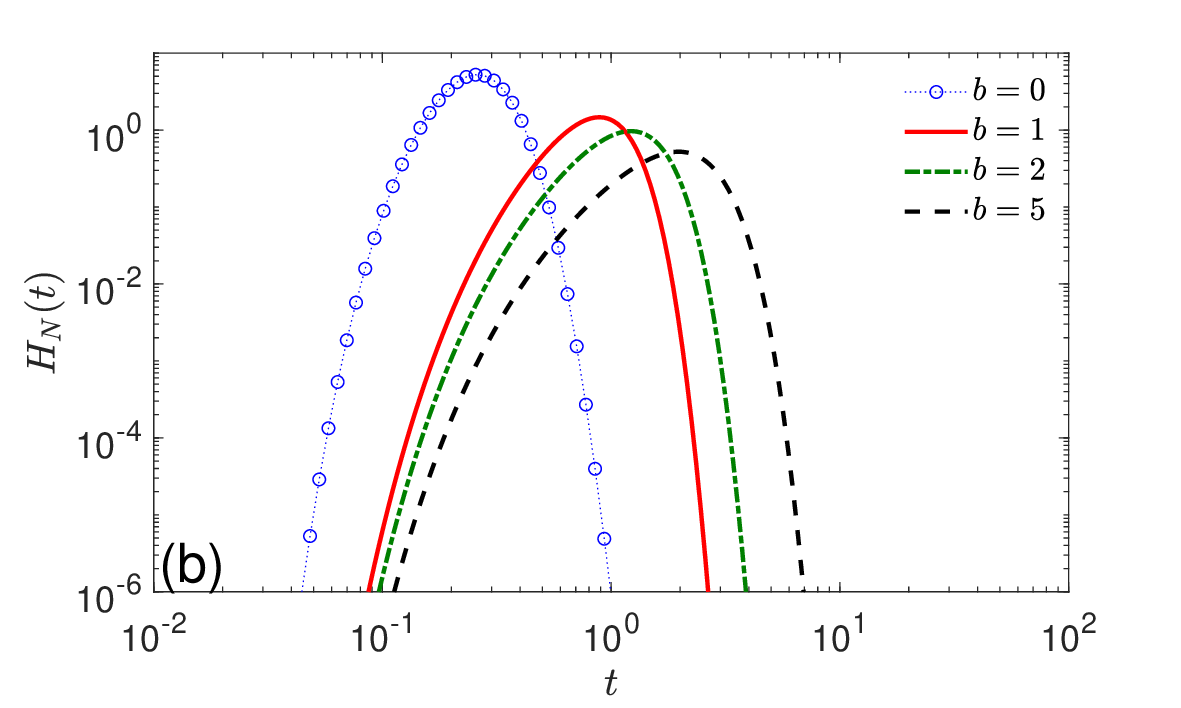} 
\end{center}
\caption{PDF $H_N(t)$ of the fastest FPT $\T_N$ for diffusion on the half-line,
with $x_0=2$, $D=1$, and $N=100$, with an extended injection modeled by the
gamma distribution (\ref{eq:gGamma}) with $\nu=2$ and four values of $b$ (see
the legend; note that $b=0$ formally corresponds to an instantaneous injection).
{\bf (a)} full plot; {\bf (b)} zoom into the vertical axis from $10^{-6}$ to
$10^1$.}
\label{fig:H_1d_gamma_b}
\end{figure}

Figure \ref{fig:H_1d_gamma_b} shows how the PDF $H_N(t)$ is affected by the
timescale $b$ of the gamma distribution (\ref{eq:gGamma}) of the entrance
times. In panel (a), one can see that both the left and the right tails of
the PDF are independent of $b$ in the limits $t\to0$ and $t\to\infty$.
However, all curves start to follow the same asymptotic behavior at extremely
small amplitudes ($\sim10^{-100}$), which are totally irrelevant for
applications. In turn, the relevant range of the amplitudes can be obtained
by zooming into this plot, as shown in panel (b). Expectedly, an increase of
$b$ implies longer entrance times and thus shifts the PDF to the right.

\section{Analysis of the most probable time}

\begin{figure}
\includegraphics[width=85mm]{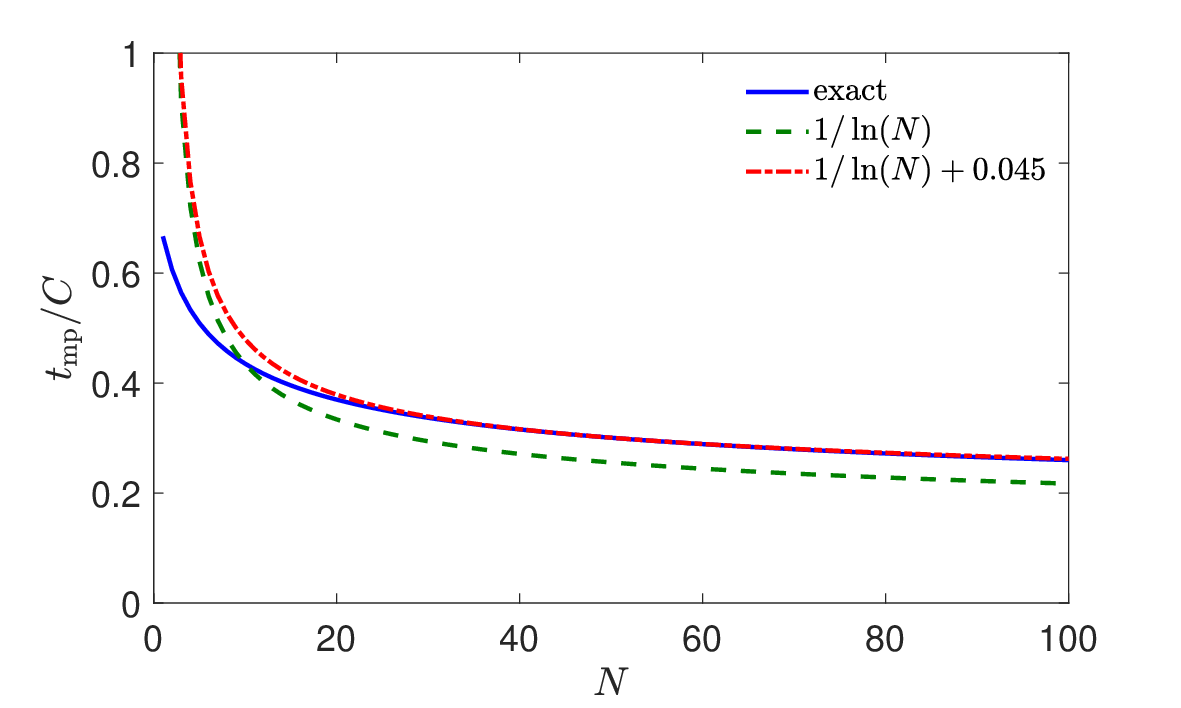} 
\caption{Most probable time $T_{\rm mp}$ (divided by $C=x_0^2/(4D)$) for
diffusion on the half-line with an instantaneous injection of $N$
particles.  The solid line represents the exact value obtained by
solving numerically Eq.~(\ref{eq:tmp_1d}); the dashed and dash-dotted
lines show two approximate asymptotic relations, $1/\ln(N)$ and
$1/\ln(N)+0.045$, the second one being shifted for a better
agreement.}
\label{fig:Tmode_1d}
\end{figure}

The most probable time $T_{\rm mp}$ characterizes the maximum of the PDF and
can thus be found by solving the equation
\begin{equation}
\partial_t H_N(t)=N[S(t)]^{N-2}\biggl(S(t)\partial_tH(t)-(N-1)[H(t)]^2\biggr)=0,
\end{equation}
which leads to
\begin{equation}
\partial_t\ln(H(t))=(N-1)\frac{H(t)}{S(t)}.
\end{equation}
For diffusion on the half-line, we substitute expressions for $H(t)$ and $S(t)$
to find
\begin{equation}
N-1=f(\xi),
\end{equation}
where $\xi=t/C$, $C=x_0^2/(4D)$ and
\begin{equation}
\label{eq:tmp_1d}
f(\xi)=\frac{\sqrt{\pi}\erf(\sqrt{1/\xi})(1-\frac{3}{2}\xi)}{\sqrt{\xi}}e^{1/
\xi}.
\end{equation}
One can check that $f(\xi)$ is a monotonously decreasing function of $\xi$
which diverges as $\xi\to0$ and vanishes at $\xi=2/3$. In a first
approximation, one can neglect slowly varying functions (as compared to $e^{
1/\xi}$) to get $N-1\approx\sqrt{\pi}e^{1/\xi}$, from which
\begin{equation}
\xi\approx\frac{1}{\ln((N-1)/\sqrt{\pi})}.
\end{equation}
Higher-order corrections (accounting for $1/\sqrt{\xi}$) are needed to
get the correct asymptotic behavior.  Nevertheless, we see that, in
our rough approximation,
\begin{equation}
T_{\rm mp}\approx\frac{C}{\ln N}=\frac{x_0^2}{4D\ln N},
\end{equation}
i.e., the most probable time behaves similarly as the mean FPT. 
This approximation can be further improved, as illustrated in
Fig.~\ref{fig:Tmode_1d}.

\end{document}